\documentclass[aps,twocolumn,amssymb,amsfonts,amsmath,showpacs,final,prb,superscriptaddress]{revtex4-2}

\usepackage{epsfig}
\usepackage{float}
\usepackage{graphicx}
\usepackage{color}
\usepackage{amsmath}
\usepackage{siunitx}
\usepackage{multirow}

\begin{document}

\title{Resonant molecular transitions in second harmonic generation spectroscopy of Fe-octaethylporphyrin adsorbed on Cu(001)} 

\author{A. Eschenlohr}
\email[]{andrea.eschenlohr@uni-due.de}
\affiliation{Faculty of Physics and Center for Nanointegration Duisburg-Essen (CENIDE), University of Duisburg-Essen, Lotharstr.~1, 47057 Duisburg, Germany}

\author{R. Shi}
\affiliation{Department of Physics, Rheinland-Pf\"alzische Technische Universit\"at, Box 3049, 67653 Kaiserslautern, Germany}

\author{J. Chen}
\affiliation{Faculty of Physics and Center for Nanointegration Duisburg-Essen (CENIDE), University of Duisburg-Essen, Lotharstr.~1, 47057 Duisburg, Germany}

\author{P. Zhou}
\affiliation{Faculty of Physics and Center for Nanointegration Duisburg-Essen (CENIDE), University of Duisburg-Essen, Lotharstr.~1, 47057 Duisburg, Germany}

\author{U. Bovensiepen}
\affiliation{Faculty of Physics and Center for Nanointegration Duisburg-Essen (CENIDE), University of Duisburg-Essen, Lotharstr.~1, 47057 Duisburg, Germany}

\author{W. H\"ubner}
\affiliation{Department of Physics, Rheinland-Pf\"alzische Technische Universit\"at, Box 3049, 67653 Kaiserslautern, Germany}

\author{G. Lefkidis}
\email[]{lefkidis@rptu.de}
\affiliation{Department of Physics, Rheinland-Pf\"alzische Technische Universit\"at, Box 3049, 67653 Kaiserslautern, Germany}
\affiliation{School of Mechanics, Civil Engineering and Architecture, Northwestern Polytechnical University, Xi'an 710072, China}

\date{\today}

\begin{abstract}

Metal-organic molecular adsorbates on metallic surfaces offer the potential to both generate materials for future (spin-)electronics applications as well as a better fundamental understanding of molecule-substrate interaction, provided that the electronic properties of such interfaces can be analyzed and/or manipulated in a targeted manner. To investigate electronic interactions at such interfaces, we measure optical second harmonic generation (SHG) from iron-octaethylporphyrin (FeOEP) adsorbed on Cu(001), and perform electronic structure calculations using coupled cluster methods including optical excitations. We find that the SHG response of FeOEP/Cu(001) is modified at 2.15-2.35~eV fundamental photon energy compared to the bare Cu(001) surface. Our polarization-dependent analysis shows that the $\chi_{zzz}^{(2)}$ non-linear susceptibility tensor element dominates this modification. The first-principles calculations confirm this effect and conclude a resonantly enhanced SHG by molecular transitions at $\hbar\omega \geq 2$~eV. We show that the enhancement of $\chi^{(2)}_{zzz}$ results from a strong charge-transfer character of the molecule-substrate interaction. Our findings demonstrate the suitability of surface SHG for the characterization of such interfaces and the potential to employ it for time-resolved SHG experiments on optically induced electronic dynamics. 

\end{abstract}

\maketitle

\section{Introduction}

\begin{figure}[t]
\includegraphics[width=0.95\columnwidth]{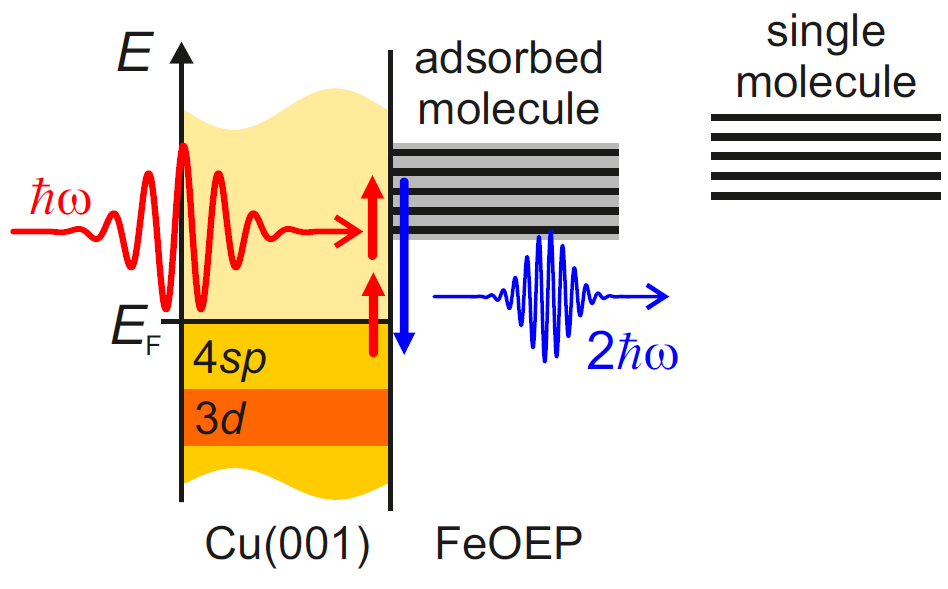}
\caption{Schematic depiction of molecular states of iron-octaethylporphyrin (FeOEP) upon adsorption at a Cu(001) surface and second harmonic generation (SHG) probe: The states of a single molecule in the gas phase shift, broaden and overlap upon adsorption but retain their molecular character. SHG at the FeOEP/Cu(001) interface probes, depending on the fundamental photon energy $\hbar\omega$ respectively the second harmonic photon energy $2\hbar\omega$, the Cu 3$d$ or $4s p$ states near the Fermi level $E_\mathrm{F}$, the FeOEP states, or, as illustrated here, a transition involving both molecule and substrate states. } 
    \label{fig1}
\end{figure}

In today's heavily digitized world, the quest for fast and miniaturized computer-logic elements becomes increasingly important. One promising path is nanospintronics, i.e., devices in which the spin rather than the charge of the electrons functions as information carrier \cite{Coronado2009,Xianrong,Gillmeister,Leiva,Bo,Zizhao,Berakdar2022,Ahn2020, Lu2021, Liu2021, Hirohata2014, Liu2020}. Spintronic devices utilize the combination of different spin states in order to build different quantum-logic gates \cite{Chaudhuri2017, Hubner2009, Wu2004, Shi2018}. One of the possible relevant concepts is the ultrafast optical manipulation of the magnetization state of magnetic systems \cite{Zhang2021, Wang2020, Panda2021, Zhang2018, Carva2017, Baibich1988, Yao2018,JPCM-review-2024}. 

Metal-organic molecules with a permanent magnetic moment adsorbed on metal surfaces are a potential realization of such future (spin-) electronics applications. In order to efficiently design future magnetic nanologic elements it is imperative to further elucidate the behavior of magnetic molecules as they interact with their immediate chemical and physical environment. One particular example of such an environmental influence, which motivates the current study, is the coupling of the magnetic molecule to the surface. In fact, the surface, rather than being inert and simply ``perturbing'' the magnetic molecule, also actively participates in its functionalization, by means of geometry stabilization, ordering in arrays, and perhaps most importantly by interconnecting several molecules into larger integrated arrays \cite{Behin-Aein2010-do,doi:10.1126/science.1201725,C5TC03225C}. 

Fig.~\ref{fig1} sketches the electronic states of a molecule/metal interface. When a molecule is adsorbed at the surface, the discrete molecular states of a previously non-interacting, isolated molecule may broaden and/or shift due to interaction with the continuum of electronic states in the metal \cite{Lindstrom2006}. The spectroscopic challenge in the analysis of such interfaces is achieving sensitivity to the interface electronic states while dealing with both discrete molecular and energetically broad metal states. Hybridization of the molecule with the metal implies an increased spatial extension of the electronic states along the surface normal which reduces the spectral weight at the interface \cite{Fauster2012}. 

We choose here optical second harmonic generation (SHG) spectroscopy for the analysis of a model molecule/metal interface, as shown schematically in Fig.~\ref{fig1}. Since the second-order non-linear susceptibility vanishes under spatial inversion symmetry, SHG is a surface- and interface-sensitive probe for centrosymmetric systems including cubic crystals \cite{Lupke1994, Bisio2009}, such as Cu(001) employed as a substrate here. SHG spectroscopy has already been applied for the characterization of molecular films in the 1-100~nm thickness range on solid substrates in earlier work, primarily phtalocyanines or porphyrins prepared on insulators or semiconductors \cite{Kumagai1993, Hoshi1995, Hoshi1996, Yamada1996, Echevarria2003, Pandey2016}. As SHG is experimentally realized with ultrashort, i.e.\ few 10~fs laser pulses, it also offers the potential for a future analysis of the ultrafast dynamics at molecule/metal interfaces in pump-probe experiments, as previously realized for ferromagnet/metal interfaces \cite{Chen2017, Chen2019} and molecular films \cite{Jailaubekov2013, Schulze2014, Hansel2017, Wirsing2019, Sivanesan2024}. 

Iron-porphyrin, which is our system of choice, has a divalent iron atom (Fe$^\mathrm{II}$) and plays an important role in electrocatalysis, environmental protection, biology and computer science \cite{Xie2021, Hillebrands2002, Sa2016}. Because the Fe$^\mathrm{II}$ atom has six $d$-electrons ([Ar]$3d^6$ configuration), depending on the exact conditions, the Fe$^\mathrm{II}$-porphyrin molecule may take a high spin state ($S=2$, quintet), intermediate spin state ($S=1$, triplet), or low spin state ($S=0$, singlet) \cite{Liao2002}.
If the central Fe atom is sixfold coordinated it can become trivalent ([Ar]$3d^5$ configuration) and, depending on the crystal field splitting of the additional ligands, the total spin can even undergo spin-crossover between sextets ($S=\frac52$) and doublets ($S=\frac12$) \cite{C6DT03859J,Engbers2023}. In fact, if the porphyrin complex is brought onto a surface, then one coordination number is saturated by the surface. In our system the extra ligand is the chlorine atom, which additionally stabilizes the whole structure while keeping the whole molecule neutral. We adopt the geometry, including the Cu(001) substrate, used by Miguel \emph{et al.} \cite{Miguel2011}, who demonstrated the magnetic interplay between the surface and the iron-porphyrin. 
Manni and Alavi showed that, due to the charge transfer excitation between the $\pi$ system of macrocycle and central metal atom, the delocalization of the electrons can stabilize the triplet spin state, where the metal atom is out of the plane of macrocycle \cite{Manni2018}. Herper \emph{et al.} used density functional theory (DFT) and x-ray absorption spectroscopy to show that a chlorine (Cl) ligand bound to the iron atom can stabilize the structure of the FeOEP molecule \cite{Herper2013}. A general practical advantage of this system is that the metal-porphyrin substance is also easily handled in experiment \cite{Liu2017}. 

In the present study, we investigate an iron-octaethylporphyrin (FeOEP) complex deposited on a metallic Cu(001) surface. More specifically, we look into the electronic interactions between the complex and the surface.
An important reason for choosing the (001) rather than the (111) surface, is  because in infinitely extended systems the band gap of the first one at the $\Gamma$ point is much larger than that of the latter one \cite{Weinelt2002,Goldmann1995}. 
To this end we experimentally and theoretically obtain non-linear optical spectra, which give insight into the nature of the surface-molecule charge- and spin-transfer states, the existence of localized surface states, and the influence of the substrate to the electronic and magnetic structure of the molecular magnet. The added value of this combined experimental and theoretical investigation lies in the quantum-chemical analysis of the charge-transfer mechanisms observed by the second-harmonic generation spectra. One important aspect of our long-term vision, namely the design and development of spintronic applications with magnetic molecules, is their integration in larger ordered matrices, which can perform complex logical calculations. Here a crucial ingredient is the coupling of the magnetic clusters to the connecting/bridging substrate, which can be metallic \cite{JPD2008-chains,Pal2009,Pal2010,PhysRevB.79.180413} or non-metallic (e.g.\ carbon-based materials \cite{Co4GNF,carbonCross,Ni3C63H54,quantumBitsNanoflakes,polyacene,JPD-graphyne}), as long as it provides delocalized electronic surface states, to which the adsorbate can couple. 

Section~\ref{exp} introduces the experimental details, i.e.\ the sample preparation and SHG spectroscopy setup, followed by the polarization analysis of the SHG signals. Then, Section~\ref{theory} describes the calculation of the FeOEP electronic states and the non-linear spectra, followed by a comparison and discussion of the experimental and theoretical results in Section~\ref{compar}. Finally, our conclusions are summarized in Section~\ref{summary}.

\section{Experiment}
\label{exp}

\subsection{Sample preparation}

\begin{figure}[t]
	\includegraphics[width=0.8\columnwidth]{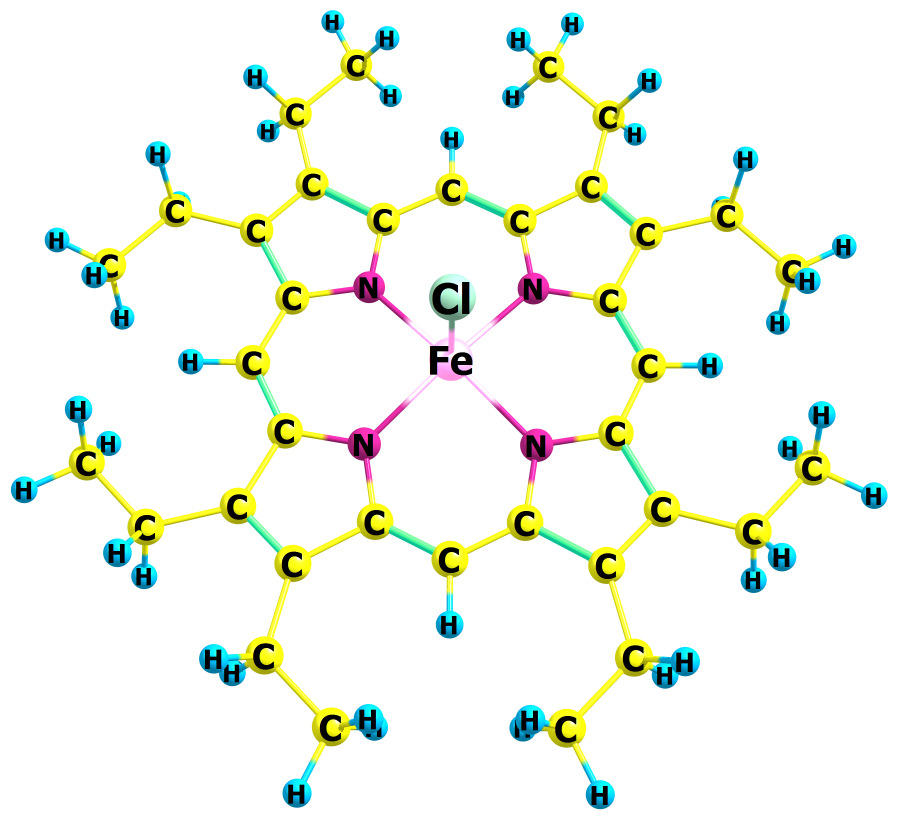}
	\caption{Optimized structure of the iron octaethylporphyrin (FeOEP) molecule. The central Fe atom lies slightly above the porphyrin plane. The stoichiometry of the complex is FeC$_{36}$H$_{44}$N$_4$Cl, and the whole complex is kept neutral, thus leading to a Fe$^{\text{III}}$ configuration. See the main text as to why we discuss a fivefold coordination for the bare molecule. The Cl ligand is oriented perpendicular to this plane. Color code: iron, pink; chlorine, green; nitrogen, red; carbon, yellow; hydrogen, blue. } 
	\label{fig:structure}
\end{figure}

FeOEP/Cu(001) interfaces are prepared and analyzed \textit{in situ} in ultrahigh vacuum at a base pressure $p<10^{-10}$~mbar at room temperature. Before deposition of the molecular adsorbate, the Cu(001) single crystal substrate is prepared by several cycles of argon ion sputtering at 1.5~keV kinetic ion energy followed by annealing to 300~°C (573~K). FeOEP is then sublimated at 212~°C (485~K) from commercially available powder purchased from Porphyrin Systems, with the layer thickness being estimated through monitoring with a quartz microbalance mounted at the evaporator outlet following the parameters established in \cite{Wende2007}. 

The structure of the iron-octaethylporphyrin (FeOEP) molecule is shown in Fig.~\ref{fig:structure}. It adsorbs flat on the Cu(001) surface \cite{Herper2013}, as schematically depicted in Fig.~\ref{fig2}(a). The chlorine ligand, which is pointing out of the plane of the molecule (see Fig.~\ref{fig:structure}) partially detaches upon adsorption on the Cu(001) surface, resulting in a ratio of approximately 1:3 for molecules with and without Cl-ligands on the surface \cite{Herper2013}. Note that in the theoretical calculations (described in Section~\ref{theory} below), which are performed on a single molecule, the Cl atom is kept. The reason for doing so is that in a centrosymmetric molecule SHG is forbidden, and thus the extra ligand provides a necessary symmetry-lowering mechanism (it turns out that the out-of-plane positioning of the central Fe atom is not enough to achieve SHG). In order to clearly identify spectral signatures belonging to the molecular adsorbate and separate them from substrate contributions, we also perform measurements of the bare Cu(001) substrate as a reference. 

\subsection{Experimental setup}
\label{exp_setup}

\begin{figure}[t]
\includegraphics[width=0.9\columnwidth]{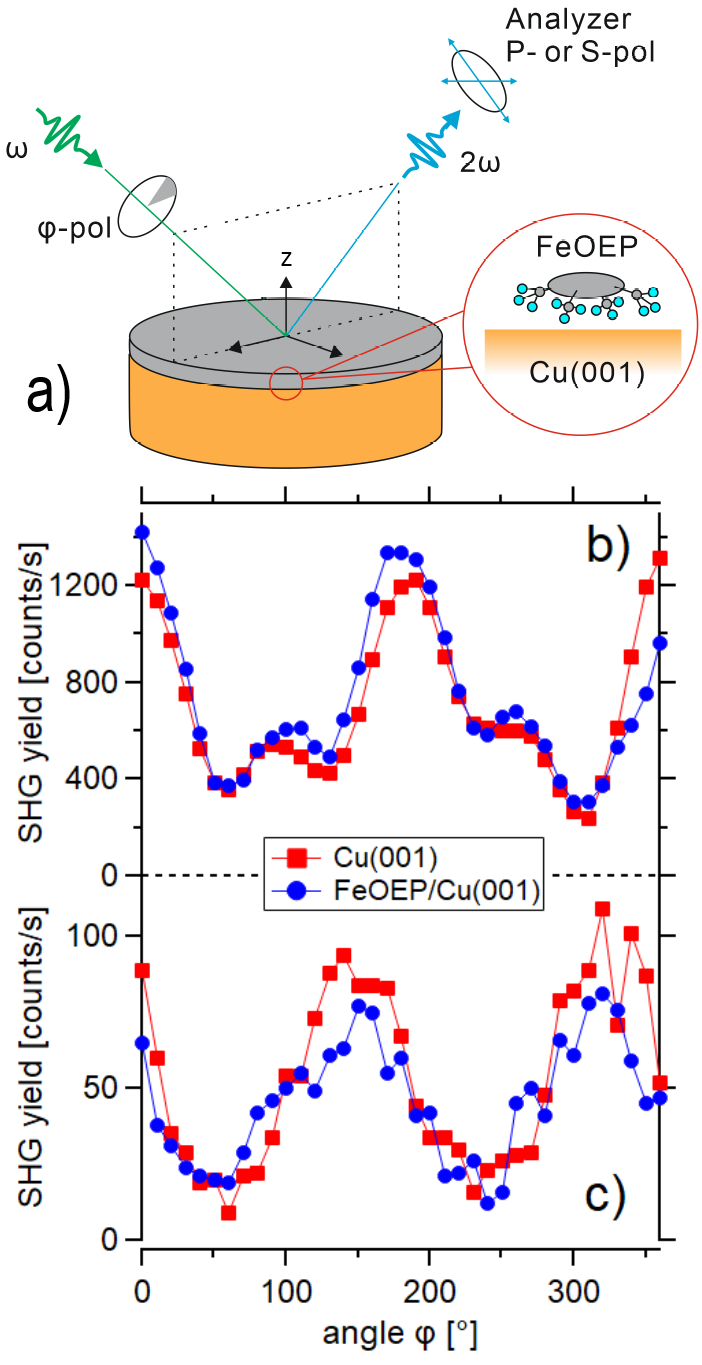}
\caption{(a) Experimental geometry: The fundamental beam with frequency $\omega$ impinges on the FeOEP/Cu(001) interface in off-normal incidence and the second harmonic at $2\omega$ is detected in reflection. While the polarization angle $\varphi$ of the fundamental beam is rotated, either P- or S-polarized SHG is analyzed. (b) P- and (c) S-polarized SHG yield depending on the polarization angle $\varphi$ of the fundamental beam with 567~nm wavelength (2.19~eV photon energy), for the bare Cu(001) substrate (red squares) and the FeOEP/Cu(001) interface (blue circles). }
    \label{fig2}
\end{figure}

The experimental geometry is depicted schematically in Fig.~\ref{fig2}(a). Ultrashort laser pulses in the visible wavelength range (1.9~eV to 2.5~eV photon energy, 20-35~fs pulse duration depending on the photon energy) are employed to generate second harmonic radiation from the FeOEP/Cu(001) interface respectively a Cu(001) surface. The ultrashort laser pulses are generated with a non-collinear optical parametric amplifier (NOPA) \cite{Wilhelm1997}, which is driven by a Ti:Sapphire regenerative amplifier operating at 100~kHz repetition rate. The laser pulses at fundamental photon energy $\hbar\omega$ impinge on the surface under an angle of $39^{\circ}$ with respect to the surface normal. The polarization angle $\varphi$ of the fundamental beam can be rotated by means of a half-wave plate. Particular polarization orientations are p- respectively s-polarization, referring to the electric field vector oriented parallel respectively perpendicular to the optical plane. The polarization of the second harmonic beam at photon energy $2\hbar\omega$ reflected from the surface is analyzed with a Glan-Taylor polarizer, which selects the detected polarization component either parallel (P-polarization) or perpendicular (S-polarization) to the optical plane. Here and in the following, capital letters indicate the polarization direction of the detected second harmonic radiation. In order to avoid accidental detection of reflected fundamental radiation, the $\hbar\omega$ beam is split off from the second harmonic beam with a wedged prism. The latter is further transmitted through a UV filter and monochromator to suppress fundamental light, before being detected by a photomultiplier tube. Single photon counting ensures the necessary high sensitivity of the setup to the weak surface SHG signal. Since the pulse energy generated by the NOPA varies with the chosen fundamental photon energy, we employ a reference measurement, namely SHG from a y-cut quartz crystal \cite{SchmittPhD}, for normalization of the SHG signal from the sample whenever we compare different photon energies. 

\subsection{Polarization analysis}
\label{exp_pol}

We first analyze the polarization dependence of the second harmonic response of the FeOEP/Cu(001) surface. The incoming polarization at frequency $\omega$ is rotated by means of a half-wave plate from p-polarized at a polarization angle of $\varphi = 0^{\circ}$ to s-polarized at $\varphi = 90^{\circ}$, while either the P- or the S-polarized component of the second harmonic radiation at frequency $2\omega$ generated by the surface is analyzed, see Fig.~\ref{fig2}(a). The measured polarization dependence at $\hbar\omega = 2.19~eV$ is shown in Fig.~\ref{fig2}(b-c) for one monolayer (ML) of FeOEP on Cu(001) and the bare metal surface. Overall, we find that the S-polarized SHG yield is about one order of magnitude smaller than the P-polarized SHG yield. As can be seen in Fig.~\ref{fig2}(b), the P-polarized SHG response exhibits two maxima at p- and s-polarization of the fundamental beam, with the yield being a factor of approximately 3 larger for the p-P compared to the s-P polarization combination. The S-polarized SHG in Fig.~\ref{fig2}(c) shows maxima of similar yield near $\varphi \approx 140^{\circ}$ and $\varphi \approx 320^{\circ}$, meaning that here a S-polarized SHG response occurs only for intermediate (or mixed) incoming polarization. 

The polarization dependence of the SHG response of the FeOEP/Cu(001) interface is very similar to that of the bare Cu(001) surface at this photon energy, see Fig.~\ref{fig2}(b-c), with only small deviations in terms of the magnitude of the SHG yield. Due to this similarity, we employ the simplified expressions for the second harmonic fields $E^{(2)}$ radiated from a (001) surface \cite{Sipe1987} 
\begin{eqnarray}
	E^{(2)}_{\text{p}-\text{P}} &=& |\gamma_{zzz}\chi_{zzz}^{(2)} + \gamma_{zxx}\chi_{zxx}^{(2)} + \gamma_{xzx}\chi_{xzx}^{(2)}|, \label{shfields-1}\\
	E^{(2)}_{\text{s}-\text{P}} &=& |\gamma_{zxx}\chi_{zxx}^{(2)}|, \label{shfields-2}\\
	E^{(2)}_{\text{mix}-\text{S}} &=& |\gamma_{xzx}\chi_{xzx}^{(2)}|, 
	\label{shfields}
\end{eqnarray}
with linear optical coefficients $\gamma_{ijk}$ and second harmonic susceptibility tensor elements $\chi_{ijk}^{(2)}$. In the employed coordinate system, the $z$-axis points in the direction of the surface normal, see Fig.~\ref{fig2}(a). Following this framework, we understand the larger p-P compared to s-P SHG yield, see Fig.~\ref{fig2}(b), as a result of a larger number of second harmonic susceptibility tensor elements being contained in $E^{(2)}_{\text{p}-\text{P}}$, combined with large absolute value of $\chi_{zzz}^{(2)}$, which describes the induced non-linear polarization along the direction of spatial symmetry breaking, i.e. normal to the sample surface. 

In the following, first principles calculations of the SHG response of FeOEP and the FeOEP/Cu(001) interface are described, before moving to a direct comparison between calculated and measured SHG spectra.

\section{Theory}
\label{theory}

\subsection{Local symmetry of Fe}

Strictly speaking the bare porphyrin molecule belongs to the C$_{\text{4v}}$ point symmetry group, since it has a fourfold $C_4$ rotation axis and no reflection axis perpendicular to it. Nonetheless, the symmetry is only slightly broken from an initial D$_{\text{4h}}$ point symmetry group. This fact, as we will see later, becomes obvious in the shape of the electronic states. For this reason, although we will be referring to the irreducible representations (irreps) of the actual symmetry, occasionally we will give in parenthesis the relevant irreps of the higher symmetry, whenever this is useful. The main reason for doing so, is that although the usual optical-selection rule which forbids electronic \emph{gerade}$\longleftrightarrow$\emph{gerade} and \emph{ungerade}$\longleftrightarrow$\emph{ungerade} transitions in D$_{\text{4h}}$ (Laporte rule) is no longer valid in the lower C$_{\text{4v}}$, the transition still remains weak, since the symmetry breaking is only small. Table \ref{tab:irreps_splitting} gives the splitting of the states of angular momentum up to $L=4$ (the highest total orbital angular momentum that can be reached with four unpaired electrons in the $d$ shell) starting from the atom (spherical symmetry K) down to D$_{\text{4h}}$, C$_{\text{4v}}$ and C$_{\text{2v}}$.

In the bare porphyrin molecule, transitions with linearly polarized light in the z direction, belonging to the irrep A$_1$ (A$_{2\text{u}}$) are very weak, because a charge redistribution in the z direction is practically impossible. This changes dramatically when the molecule is deposited on a surface, since charge-transfer excitations between adsorbate and substrate are possible.

\begin{table}[t]{
\caption{Splitting due to local crystal field splitting, from the atom (spherical symmetry K), down to D$_{\text{4h}}$, C$_{\text{4v}}$ and C$_{\text{2v}}$, for orbital momentum up to $L=4$ (G states). The quantum chemical calculations are performed in C$_{\text{2v}}$, which is the largest Abelian subgroup of D$_{\text{4h}}$.
    \label{tab:irreps_splitting}}
    \centering
    \begin{eqnarray*}
        \begin{matrix}
            \text{K}&&\text{D}_\text{4h}&&\text{C}_\text{4v}&&\text{C}_\text{2v}\\
            \hline\hline
            \text{S}&\longrightarrow&\text{A}_\text{1g}&\begin{matrix}\longrightarrow\end{matrix}&\text{A}_\text{1}&\begin{matrix}\longrightarrow\end{matrix}&\text{A}_\text{1}
            \\
            \hline
            \text{P}&\longrightarrow& \left\{\begin{matrix}\text{A}_\text{2u}\\\text{E}_\text{u}\end{matrix}\right.&\begin{matrix}\longrightarrow\\\longrightarrow\end{matrix}&\begin{matrix}\text{A}_\text{1}\\\text{E}\end{matrix}&\begin{matrix}\longrightarrow\\\longrightarrow\end{matrix}&\begin{matrix}\text{A}_\text{1}\\\text{B}_1\oplus \text{B}_2\end{matrix}
            \\
            \hline
            \text{D}&\longrightarrow& \left\{\begin{matrix}\text{A}_\text{1g}\\\text{B}_\text{1g}\\\text{B}_\text{2g}\\\text{E}_\text{g}\end{matrix}\right.&
            \begin{matrix}\longrightarrow\\\longrightarrow\\\longrightarrow\\\longrightarrow\end{matrix}&\begin{matrix}\text{A}_\text{1}\\\text{B}_\text{2}\\\text{B}_\text{1}\\\text{E}\end{matrix}
            &
            \begin{matrix}\longrightarrow\\\longrightarrow\\\longrightarrow\\\longrightarrow\end{matrix}&\begin{matrix}\text{A}_\text{1}\\\text{A}_\text{2}\\\text{A}_\text{1}\\\text{B}_1\oplus \text{B}_2\end{matrix}
            \\\hline
            
            \text{F}&\longrightarrow& \left\{\begin{matrix}\text{A}_\text{2u}\\\text{B}_\text{1u}\\\text{B}_\text{2u}\\\text{2E}_\text{u}\end{matrix}\right.&
            \begin{matrix}\longrightarrow\\\longrightarrow\\\longrightarrow\\\longrightarrow\end{matrix}&\begin{matrix}\text{A}_\text{1}\\\text{B}_\text{2}\\\text{B}_\text{1}\\\text{2E}\end{matrix}
            &
            \begin{matrix}\longrightarrow\\\longrightarrow\\\longrightarrow\\\longrightarrow\end{matrix}&\begin{matrix}\text{A}_\text{1}\\\text{A}_\text{2}\\\text{A}_\text{1}\\\text{2B}_1\oplus \text{2B}_2\end{matrix}
            \\\hline
            
            \text{G}&\longrightarrow& \left\{\begin{matrix}\text{2A}_\text{1g}\\\text{A}_\text{2g}\\\text{B}_\text{1g}\\\text{B}_\text{2g}\\\text{2E}_\text{g}\end{matrix}\right.&
            \begin{matrix}\longrightarrow\\\longrightarrow\\\longrightarrow\\\longrightarrow\\\longrightarrow\end{matrix}&\begin{matrix}\text{2A}_\text{1}\\\text{A}_\text{2}\\\text{B}_\text{1}\\\text{B}_\text{2}\\\text{2E}\end{matrix}
            &
            \begin{matrix}\longrightarrow\\\longrightarrow\\\longrightarrow\\\longrightarrow\\\longrightarrow\end{matrix}&\begin{matrix}\text{2A}_\text{1}\\\text{A}_\text{2}\\\text{A}_\text{1}\\\text{A}_\text{2}\\\text{2B}_1\oplus \text{2B}_2\end{matrix}
        \end{matrix}
    \end{eqnarray*}
}
\end{table}

Mathematically, the inclusion of spin-orbit coupling (SOC) obliges us to use double point groups (C$_{\text{4v}}^*$ and D$_{\text{4h}}^*$), because our system has an odd number of electrons (we calculate doublet and quartet states), giving rise to two additional two ($\Gamma_6$, $\Gamma_7$) and four ($\Gamma_6^+$, $\Gamma_7^+$, $\Gamma_6^-$, $\Gamma_7^-$) irreducible co-representations, respectively \cite{doubleGroups,Altmann}. There is no four-dimensional irrep, which means that quartet states are split into two doubly degenerate pairs (Koopmans' degeneracy in the absence of a Zeeman splitting). Due to the fact, however, that the central magnetic atom does not retain its full rotational symmetry, the effective gyromagnetic ratio ($g$-factor) of the orbital angular momentum shrinks dramatically, and hence the zero-field splitting ($\approx 0.454$ meV for the doublet ground state of the bare iron-porphyrin and even smaller when the surface is taken into account) is very small compared to the thermal fluctuations at room temperature. Note that although after SOC neither $\hat{\mathbf{L}}$ nor $\hat{\mathbf{S}}$ are good quantum numbers anymore, it is still possible to compute their expectation values $\langle\hat{\mathbf{L}}\rangle$ and $\langle\hat{\mathbf{S}}\rangle$. A full theoretical analysis as well as the calculation of the $g$-factor for the different spectroscopic terms necessitates magnetic-field-resolved spectra. As it goes beyond the scope of the present work it will be presented elsewhere.

The $d^6$ configuration of Fe$^\text{II}$ locally yields 16 many-body states,
namely ${^1\text{I}}\oplus{^3\text{H}}\oplus{^3\text{G}}\oplus{2^1\text{G}}\oplus{2^3\text{F}}\oplus{^1\text{F}}\oplus{^5\text{D}}\oplus{^3\text{D}}\oplus{2^1\text{D}}\oplus{2^3\text{P}}\oplus{2^1\text{S}}$  (altogether 210 substates). Here we use the usual nomenclature $^{2s+1}L$, where a capital letter refers to the total (many-electron) orbital angular momentum. If we restrict ourselves to maximally four unpaired electrons (up to G states and up to quintets) since our computational method includes up to two-electron primary virtual excitations starting from a closed shell configuration, SOC leads to the spectroscopic terms 
${^3\text{G}_5}$, ${^3\text{G}_4}$, ${^3\text{G}_3}$, 
${^1\text{G}_4}$, 
${^3\text{F}_4}$, ${^3\text{F}_3}$, ${^3\text{F}_2}$,
${^1\text{F}_3}$,
${^5\text{D}_4}$, ${^5\text{D}_3}$, ${^5\text{D}_2}$, ${^5\text{D}_1}$, ${^5\text{D}_0}$, 
${^3\text{D}_3}$, ${^3\text{D}_2}$, ${^3\text{D}_1}$, 
${^1\text{D}_2}$, 
${^3\text{P}_2}$, ${^3\text{P}_1}$, and ${^3\text{P}_0}$ (in the $^{2s+1}L_j$ nomenclature). Being of even multiplicity, all these terms can be studied using the simple point groups
C$_{\text{4v}}$ and D$_{\text{4h}}$. 
The $d^5$ configuration of Fe$^\text{III}$ locally yields 16 many-body states, namely 
${^2\text{I}}\oplus
{^2\text{H}}\oplus
{^4\text{G}}\oplus
{2^2\text{G}}\oplus
{^4\text{F}}\oplus
{2^2\text{F}}\oplus
{^4\text{D}}\oplus
{3^2\text{D}}\oplus
{^4\text{P}}\oplus
{^2\text{P}}\oplus
{^6\text{S}}\oplus
{^2\text{S}}$ (altogether 252 substates). 
Considering configurations with up to three unpaired electrons (one single-electron virtual excitation and one ionization when starting from a closed-shell reference), i.e.\ up to F states and up to quartets leads
to the spectroscopic 
${^4\text{F}_{9/2}}$, ${^4\text{F}_{7/2}}$, ${^4\text{F}_{5/2}}$, ${^4\text{F}_{3/2}}$,
${^2\text{F}_{7/2}}$, ${^2\text{F}_{5/2}}$,
${^4\text{D}_{7/2}}$, ${^4\text{D}_{5/2}}$, ${^4\text{D}_{3/2}}$, ${^4\text{D}_{1/2}}$,
${^2\text{D}_{5/2}}$, ${^2\text{D}_{3/2}}$,
${^4\text{P}_{5/2}}$, ${^4\text{P}_{3/2}}$, ${^4\text{P}_{1/2}}$,
${^2\text{P}_{3/2}}$, ${^2\text{P}_{1/2}}$, and ${^2\text{S}_{1/2}}$
terms, for the symmetry analysis of which double point groups are necessary.
Due to the fourfold symmetry of the surrounding crystal field, the Fe$^\text{II}$ states get split into doubly degenerate substates, while the Fe$^\text{III}$ states get split into doubly degenerate substates and one single non-degenerate state (see Appendix A for the character tables of the relevant double point groups).

\subsection{Electronic states}
One important aspect to consider in our system is that we are interested in describing magnetism (for which static correlations are needed), as well as optically induced electronic transitions (which are governed mainly by dynamic correlations). For this reason we use coupled-cluster (CC) methods, which not only can adequately handle both kinds of correlations \cite{JAP.109.07D303,correlations}, but also yield many-body electronic states (depending on the size of the system they can cover the energy spectrum up to a few eV). 

We first optimize the geometry of the FeOEP molecule, which belongs to the C$_{4\text{v}}$ symmetry point group (compare Fig.~\ref{fig:structure}), at the restricted open-shell Hartree-Fock (ROHF) level with the 6-31G* basis set for Fe and 3-21G for the ligand atoms. Then we employ the coupled-cluster method with single and double excitations (CCSD) to get a correlated reference many-body state \cite{Hanna2017, Purvis1982}. Subsequently, we compute the excited states with the equation-of-motion coupled cluster method with single and double excitations (EOM-CCSD) \cite{Liedy2020, Stanton1993, Kowalskia2001}. The reason of our particular choice of methods lies in the ability of the coupled-cluster methods to capture to a great extent both static correlations, which are necessary for the description of magnetic systems, and dynamic correlations, which greatly affect optical transitions \cite{Dutta2020}.  We further perform calculations of the FeOEP molecule deposited on a Cu(100) surface (two atomic layers), a method which has been proven successful in finding surface-localized states in similar systems, e.g.\ when calculating the optical absorption spectra of Pt$_2$ and Pt$_4$ clusters on a Cu(001) surface \cite{Pal2009,Pal2010}. For our system we restrict the substrate to two atomic layers, which reaches the limits of today's computational capabilities for CC methods (encompassing 137 Cu atoms). The Cu cluster used bears the same symmetry as FeOEP, so that that the optical selection rules are not artificially altered. The calculations are performed with the freely available GAMESS quantum chemical package \cite{Barca2020}.

\begin{figure}[t]
	\includegraphics[width=\columnwidth]{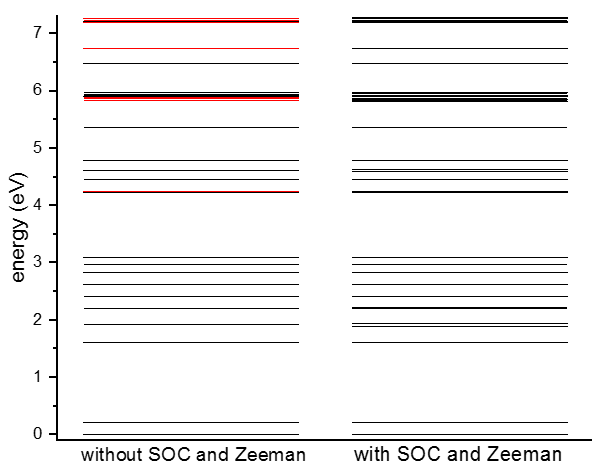}
	\caption{Excited many-body states of FeOEP for energies with respect to the ground state. (left) 40 many-body states calculated with the IP-EOM-CCSDt method, with the 30 doublets shown in black and the 10 quartets in red (cf.\ Tab.\ \ref{tab:energies}). (right) 100 states after inclusion of SOC and the Zeeman splitting with an external static magnetic field $B = 10^{-5}$~at.~un. } 
	\label{fig:level-scheme}
\end{figure}

\begin{table}[t]{
    \caption{Some many-body-state energies of the bare FeOEP before inclusion of SOC [cf.\ Fig.\ \ref{fig:level-scheme}(left)]. The upper left index in the irreps denotes the spin multiplicity. After the inclusion of SOC these irreps must be multiplied with $\Gamma_6$ for doublets and with $\Gamma_6\oplus\Gamma_7$ for quartets, which are the splittings of the representations of $S=\frac12$ and $S=\frac32$, respectively (cf.\ Tab.\ \ref{tab:characterTableC4v}).}
    \label{tab:energies}
    \centering
    \begin{tabular}{ccc|ccc}
         number&energy &irrep &number&energy  &irrep \\
         &(eV)  &(in C$_\text{4v}$) && (eV) & (in C$_\text{4v}$)\\
         \hline
          1&0.000&$^2$A$_2$ &16&5.349&$^2$E\\
          2&0.209&$^2$A$_1$ & 17&5.821&$^4$B$_2$\\
          3&1.599&$^2$E & 18&5.858&$^4$A$_1$\\
          4&1.910&$^2$E & 19&5.888&$^4$B$_2$\\
          5&2.205&$^2$B$_2$ & 20&5.900&$^2$B$_2$\\
          6&2.401&$^2$A$_1$ & 21&5.902&$^2$A$_1$\\
          7&2.601&$^2$B$_1$ & 22&5.911&$^4$B$_2$\\
          8&2.822&$^2$E & 23&5.926&$^4$A$_1$\\
          9&2.958&$^2$A$_1$ & 24&5.964&$^4$A$_1$\\
          10&3.084&$^2$E &25&6.473&$^2$E\\
          11&4.226&$^2$A$_2$ &26&6.742&$^4$E\\
          12&4.235&$^4$E& 27&7.188&$^4$B$_2$\\
          13&4.455&$^2$A$_1$ &28&7.211&$^2$B$_2$\\
          14&4.607&$^2$E &29&7.220&$^4$A$_1$\\
          15&4.781&$^2$E &30&7.263&$^4$B$_2$\\
    \end{tabular}
}
\end{table}

Subsequently, SOC and an external infinitesimally small static magnetic field are perturbatively added 
\begin{equation}
	\hat{H} =
	\sum_{i=1}^{n}\frac{Z_{a}^{\text{eff}}}{2c^{2}\mathrm{\bf
			r}_{i}^{3}}\hat{{\mathrm{\bf L}}}\cdot\hat {\mathrm{\bf
			S}}+\sum_{i=1}^{n}\mu_{\text{L}}\hat{\mathrm{\bf L}}\cdot{\bf
		B}_{\text{stat}}+\sum_{i=1}^{n}\mu_{\text{S}}\hat{\mathrm{\bf S}}\cdot{\bf
		B}_{\text{stat}},
\end{equation}
where \textbf{\^{S}} and \textbf{\^{L}} are the spin and the orbital angular momentum operators, and $\mu_{L}$ and $\mu_{S}$ are their respective magnetic moments. $Z_{a}^{\text{eff}}$ are effective nuclear charges which account for the two-electron contributions to SOC \cite{Koseki1998}, and ${\bf B}_{\text{stat}}$ is the magnetic field. The energy levels before and after the inclusion of SOC and the Zeeman interaction are depicted in Fig.~\ref{fig:level-scheme}. 

The calculation yields altogether 30 states, namely 22 doublets and 8 quartets. The computation is performed in the largest Abelian subgroup, namely C$_\text{2v}$, and the resulting states are inspected manually to decide their C$_\text{2v}$ irreps. The bare system has in total 331 electrons, giving rise to even multiplicities, (see Tab.\ \ref{tab:energies} and Fig.~\ref{fig:level-scheme}). 
Note that since we aim at distinguishing between the surface and the molecule contributions, and in order to eliminate incongruities due to different stoichiometries we also use the fivefold coordinated Fe$^{\text{III}}$ for both the bare and the surface deposited iron-porphyrine calculations. 

\begin{figure}[t]
	\includegraphics[width=\columnwidth]{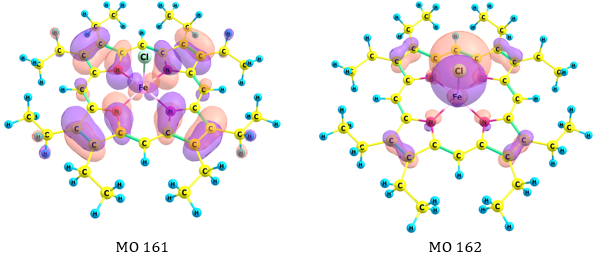}
	\includegraphics[width=\columnwidth]{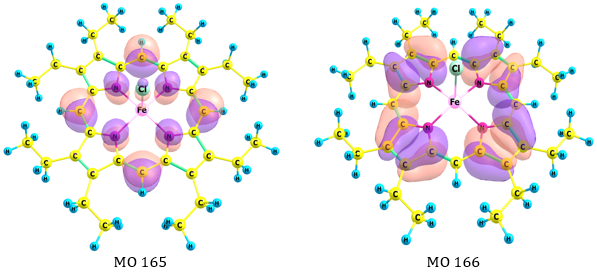}
	\includegraphics[width=\columnwidth]{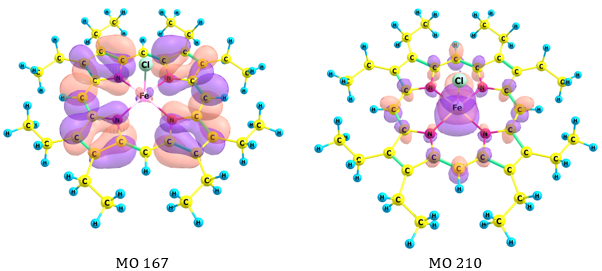}
	\caption{Selected important molecular orbitals (MOs) of FeOEP: MOs  161, 162, 165, 166, 167 and 210 have contributions both from to the Fe-$d$ orbitals and from the ligand atoms, while MO 166 is mainly localized on the ligand. MO 166 is the highest-occupied molecular orbital (HOMO) and MO 167 is the lowest unoccupied molecular orbital (LUMO). }
	\label{fig:FeOEP-MOs}
\end{figure}

In order to calculate pure-spin states the EOM-CCSD method needs to start from a closed shell Hartree-Fock determinant, which in our case carries a negative charge and has a total of 332 electrons. Then we apply the ionization-potential EOM-CCSD method (IP-EOM-CCSD), which calculates the requested 40 many-body states by removing the extra electron. Fig.~\ref{fig:FeOEP-MOs} depicts some of the most important molecular orbitals (MOs) of the FeOEP molecule. MO 166 is the highest-occupied MO (HOMO) and MO 167 is the lowest unoccupied MO (LUMO). MO 162 is an example of a P-character orbital localized on the porphyrin. These MOs participate in the virtual excitations of both the ground state and the energetically lowest electronic states. For example, one can recognize the $d_{x^2-y^2}$ character MO 155 at $-0.4093$ eV and irrep B$_1$ (B$_{1\text{g}}$), the conjugated bonding $\pi$-character of the ligand carbons in MO 159 at $-0.3303$ eV and irrep B$_2$ (B$_{1\text{u}}$), MO 161  at $-0.3164$ eV and irrep E (E$_{\text{g}}$), MO 166 at $-0.1401$ eV and irrep A$_2$ (A$_{1\text{u}}$), and MO 167 at $-0.0017$ eV and irrep E (E$_{\text{u}}$), as well as the $d_{z^2}$ character of the empty MO 210 at $0.3584$ eV and irrep E (E$_{1\text{g}}$). The many-body ground state after the inclusion of SOC practically has only one singly-occupied-MO (SOMO) contribution, namely MO 166, which is mainly a conjugated $\pi$ orbital of the four pyrrole rings of the porphyrin. 
Interestingly the highest occupied orbitals and the LUMO have all \emph{gerade} symmetry (in the higher D$_{\text{4h}}$ point group), which means that for optical transitions to occur virtual excitations from lower \emph{ungerade} MOs are needed. This is changed by the surface which  provides optically active vertical charge transitions, but only for linearly polarized light in the z direction. 

\subsection{Non-linear optical spectra}

The electric polarization in a material is a function of the incident electric field ${\bf E}$ \cite{Satitkovitchai2003}
\begin{equation}
	{\bf P}=
	\chi^{(1)}{\bf E}+\chi^{(2)}{\bf E}{\bf E}+\cdots,
\end{equation}
where $\chi^{(1)}$ and $\chi^{(2)}$ are the first- and second-order susceptibilities [please note that differently from here, $E^{(2)}$ in Eqs.\ (\ref{shfields-1})-(\ref{shfields}) refers to the emitted light]. The elements of the latter can be calculated as follows \cite{Hubner1993, Shen, Lefkidis2005, Lefkidis2006} 
\begin{widetext}
\begin{eqnarray}
	\chi_{ijk}^{(2)}\propto q\sum_{\alpha\beta\gamma}
	\Bigg[	\langle\gamma|i|\alpha\rangle\overline{\langle\alpha|j|\beta\rangle \langle\beta|k|\gamma\rangle}\,\frac{\frac{f(E_{\gamma})-f(E_{\beta})}{E_{\gamma}-E_{\beta}-\hbar\omega+i\hbar\Gamma}-
	\frac{f(E_{\beta})-f(E_{\alpha})}{E_{\beta}-E_{\alpha}-\hbar\omega+i\hbar\Gamma}}
	{E_{\gamma}-E_{\alpha}-2\hbar\omega+2i\hbar\Gamma}\Bigg],
\end{eqnarray}
\end{widetext}
where $i,j,k=x,y,z$ are components of the dipole-operator $\hat {\bf r}$. $\Gamma$ is a broadening coefficient (not explicitly calculated here), $q$ is the electron charge, 
and $E_{\alpha/\beta/\gamma}$ is the energy of state $\alpha/\beta/\gamma$. 
$\omega$ is the frequency of the fundamental (absorbed) light, while the emitted light has frequency $2 \omega$. The Boltzmann distribution 
\begin{eqnarray}f(E_{\alpha})=\frac{e^{-\frac{E_{\alpha}}{K_\text{B}T}}}{Z}
\end{eqnarray}
gives the population of the many-body state $|\alpha\rangle$ at temperature $T$ (assuming thermal equilibrium). $Z=\sum_{i=1}^{n}{e^{-\frac{E_{i}}{K_\text{B}T}}}$ is the partition function of the system and 
$K_\text{B}$ is the Boltzmann constant. The overbar denotes symmetrization with respect to the two incident photons and accounts for their indistinguishability. Due to the system's C$_\text{4V}$ symmetry $\chi^{(2)}$ contains only five non-zero elements, out of which only three are linearly independent: $\chi^{(2)}_{zxx}=\chi^{(2)}_{zyy}$, $\chi^{(2)}_{xzx}=\chi^{(2)}_{yzy}$, and $\chi^{(2)}_{zzz}$ \cite{Shen}. 

\begin{figure}[t]
	\includegraphics[width=\columnwidth]{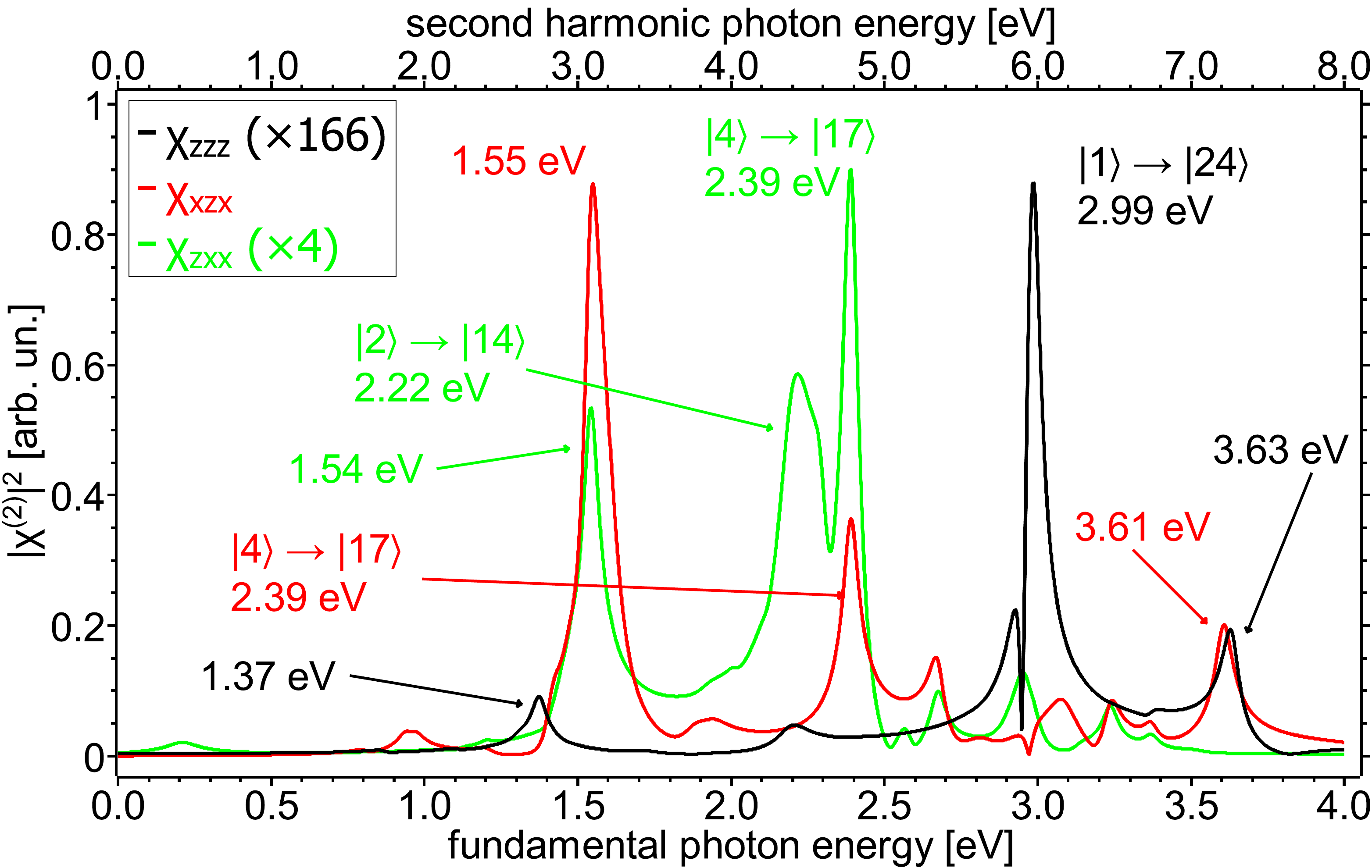}
	\caption{Theoretically calculated non-zero elements of the second harmonic susceptibility tensor, $\chi^{(2)}_{zzz}$ (166 times magnified), $\chi^{(2)}_{xzx}$, and $\chi^{(2)}_{zxx}$ (4 times magnified) for the bare FeOEP molecule in arbitrary units. Please note that energies in eV are given above the respective spectral peaks, where also the main transitions responsible for the peaks are indicated, unless the peak results from a multitude of excitations. The spectra are calculated at room temperature $T=300$ K.}
	\label{fig:theoretical-SHG}
\end{figure}

Fig.~\ref{fig:theoretical-SHG} shows the theoretically calculated non-vanishing elements for the FeOEP molecule at room temperature ($T=300$ K). Interestingly, all tensor elements have a peak at 2.2 eV or 2.4 eV, corresponding to many-body excitations $|2\rangle\!\rightarrow\!|14\rangle$ and $|4\rangle\!\rightarrow\!|17\rangle$, respectively. Note that $|2\rangle$ and $|4\rangle$ are substates of the same triplet, and therefore lie very close energetically. State $\vert 14 \rangle$ includes a virtual transition from MO 166 to MO 167 (HOMO$\rightarrow$LUMO), and an electron removed mainly from MO 159 (which becomes a SOMO) with amplitude 0.9255. The peak at 2.39 eV is due to the many-body-state transition $\vert 4 \rangle\!\rightarrow\!\vert 17 \rangle$. State $\vert 4 \rangle$  has the electron removed from MO 165 with amplitude of 0.944. State $\vert 17 \rangle$ has the electron removed from MO 155 with amplitude 0.773 and includes a virtual excitation from MO 162 to MO 210. 

The p-P signal is determined by the tensor elements $\chi^{(2)}_{zxx}$, $\chi^{(2)}_{zzz}$, and $\chi^{(2)}_{xzx}$, while the s-P signal is determined by $\chi^{(2)}_{zxx}$. In all cases the transitions along the $z$ axis play a major role, rendering the existence of the substrate extremely important. 

\begin{figure}[t]
	\includegraphics[width=\columnwidth]{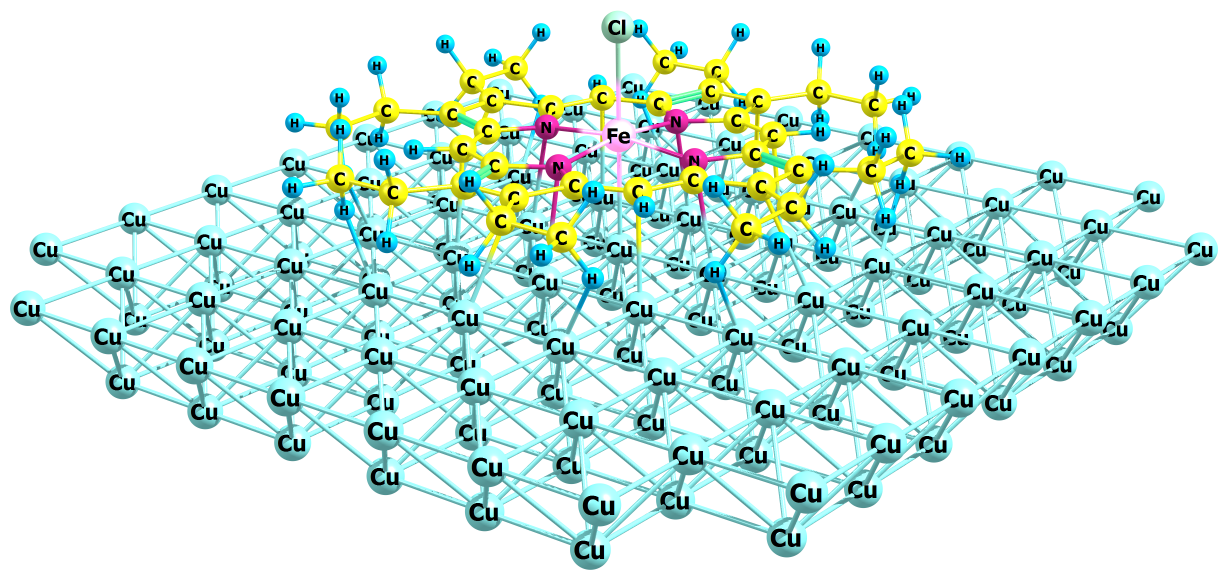}
	\caption{Structure of FeOEP adsorbed on a Cu(001) bilayer. The point group symmetry of the whole system is C$_{\text{4v}}$. Color code: iron, pink; chlorine, green; nitrogen, red; carbon, yellow; hydrogen, blue; copper, light blue. }
	\label{fig:porphyrin@Cu}
\end{figure}

\begin{figure}[t]
	\includegraphics[width=\columnwidth]{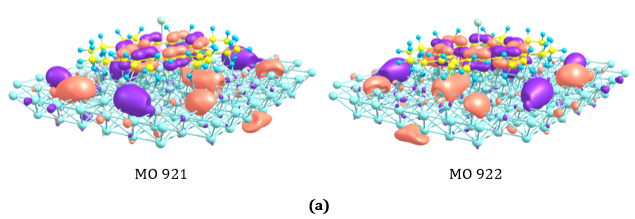}
	\includegraphics[width=\columnwidth]{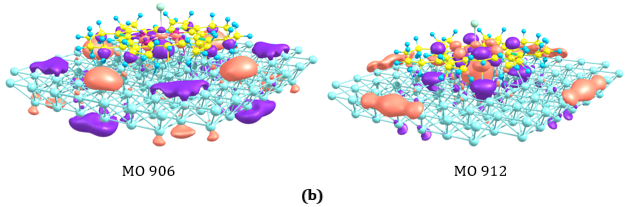}
	\includegraphics[width=\columnwidth]{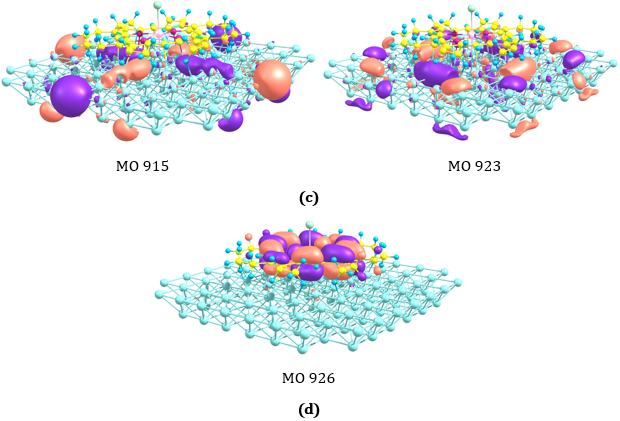}
	\caption{Selected MOs of FeOEP/Cu(001) which participate in the many-body-state transition $\vert 1 \rangle\!\rightarrow\!\vert 51 \rangle$: (a) MOs localized at the edges of the Cu bilayer that result from the specific geometry chosen, (b) porphyrin-to-surface charge transfer MOs, (c) pure Cu(001) surface MOs, (d) a pure FeOEP MO. }
	\label{fig:FeOEP-surface-MOs}
\end{figure}

In order to better elucidate the interactions between FeOEP and the substrate, we deposit the molecule on a Cu bilayer (one with 77 and a second one with 60 Cu atoms), which has the same symmetry as FeOEP, and repeat the theoretical calculations (Fig.~\ref{fig:porphyrin@Cu}). The distance between the Fe atom in FeOEP and the surface is set to 2.64 \AA\  \cite{Herper2013}. Some important MOs are depicted in Fig.~\ref{fig:FeOEP-surface-MOs}. Among them one can distinguish four categories, namely MOs which are mainly localized at the edges of the substrate (and therefore irrelevant in a more realistic/infinitely extended system), MOs which include both FeOEP and surface atomic orbitals, as well as MOs which are either pure FeOEP or pure surface states. Interestingly enough, the edge MOs lie energetically relatively deep and do not participate in the virtual excitations of the many-body states. Therefore, they do not pose a problem in our computations. 

\begin{figure}[t]
	\includegraphics[width=0.6\columnwidth]{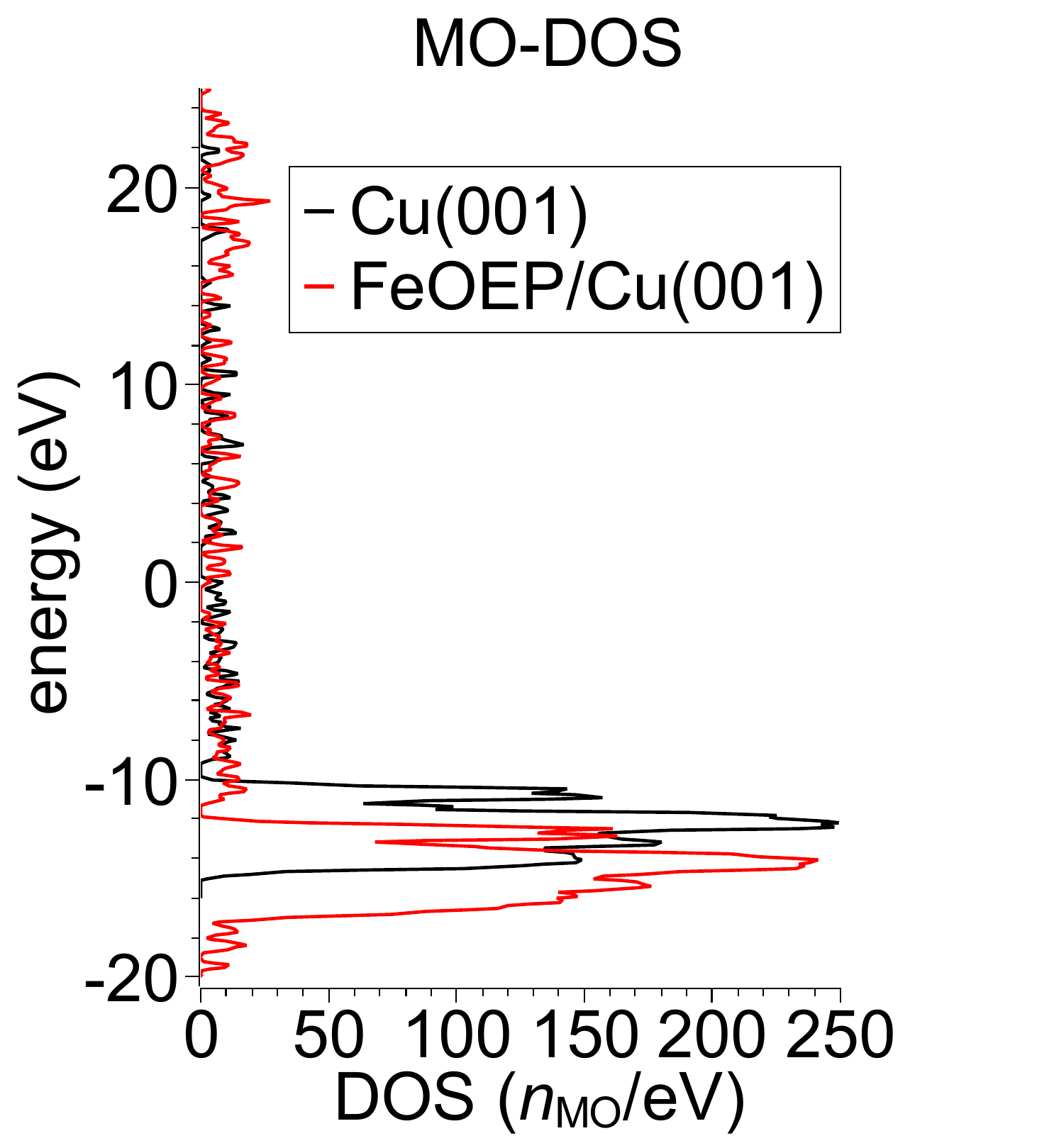}
	\caption{One-electron density-of-states (DOS) at the HF level for the bare Cu(001) surface (black line) and the combined FeOEP/Cu(001) system (red line). Only the window between -20 and 20 eV is depicted, which exhibits an appreciable density. Here the energy of the HOMO is taken as the zero energy and corresponds to the Fermi level.}
	\label{fig:DOS}
\end{figure}

Fig.~\ref{fig:DOS} depicts the electronic density-of-states (DOS) at the HF level (i.e., one-electron states before accounting for electronic correlations) of both the bare Cu(001) surface and the combined FeOEP/Cu(001) system. We see that for the combined system the shape of contribution of the surface states remains practically the same, however it sinks relatively to the Fermi level (which is set equal to the energy of the HOMO orbital) by about 2 eV. This energy shift is due to the strong interaction between the Cu surface and the porphyrin molecule. The porphyrin states do not gather in any particular energy region, but rather scatter across the whole spectrum and thus cannot be visually discerned. At the HF level, once FeOEP is deposited on Cu, charge corresponding to slightly above 0.8 electrons migrates from it into the Cu surface (also cf.\ Table \ref{tab:charge-diffusion}). This vertical charge transfer leads to more pronounced electric dipole moments along the z direction, both for the first- and the second-order susceptibility. Note also, that if the calculation included more Cu layers, then we would expect even stronger dipole moments, since the electrons would travel even further for the charge-transfer states. 

Since the addition of 137 Cu atoms increases the number of atomic orbitals, the resulting MOs and also the many-body states are more densely packed together. 
For this reason the combined FeOEP-surface calculation yields a lower-energy window (up to about 1.77 eV). Fig.~\ref{fig:theoretical-FeOEP-surface-SHG} depicts $\chi^{(2)}$ for the combined system and Cu bilayer alone. 

Note that although qualitatively our calculations indicate the importance of the FeOEP-to-surface charge transfer states, since we restrict ourselves to two Cu layers, the scattering length is inadvertently also restricted, and therefore the strength of the $\chi^{(2)}$ components which include the $z$ direction is underestimated. When comparing the theoretical and experimental SHG spectra of the combined FeOEP/Cu(001) system, the reader should also keep in mind that due to the sheer multitude of discrete states, our computations only yield many-body states up to 2.089 eV. Therefore, the calculation of the second-order susceptibility inevitably does not contain all state contributions for energies above this threshold. This is why the combined calculation does not cover the experimental range at $>4$~eV.  
Nonetheless, since we already know that the spectrum of the bare FeOEP has a peak at this area, we surmise that the presence of the surface can only intensify all peaks which contain excitations along the z direction (i.e., the $\chi_{zxx}^{(2)}$, $\chi_{xzx}^{(2)}$ and $\chi_{zzz}^{(2)}$ elements).

\begin{figure}[t]
	\includegraphics[width=\columnwidth]{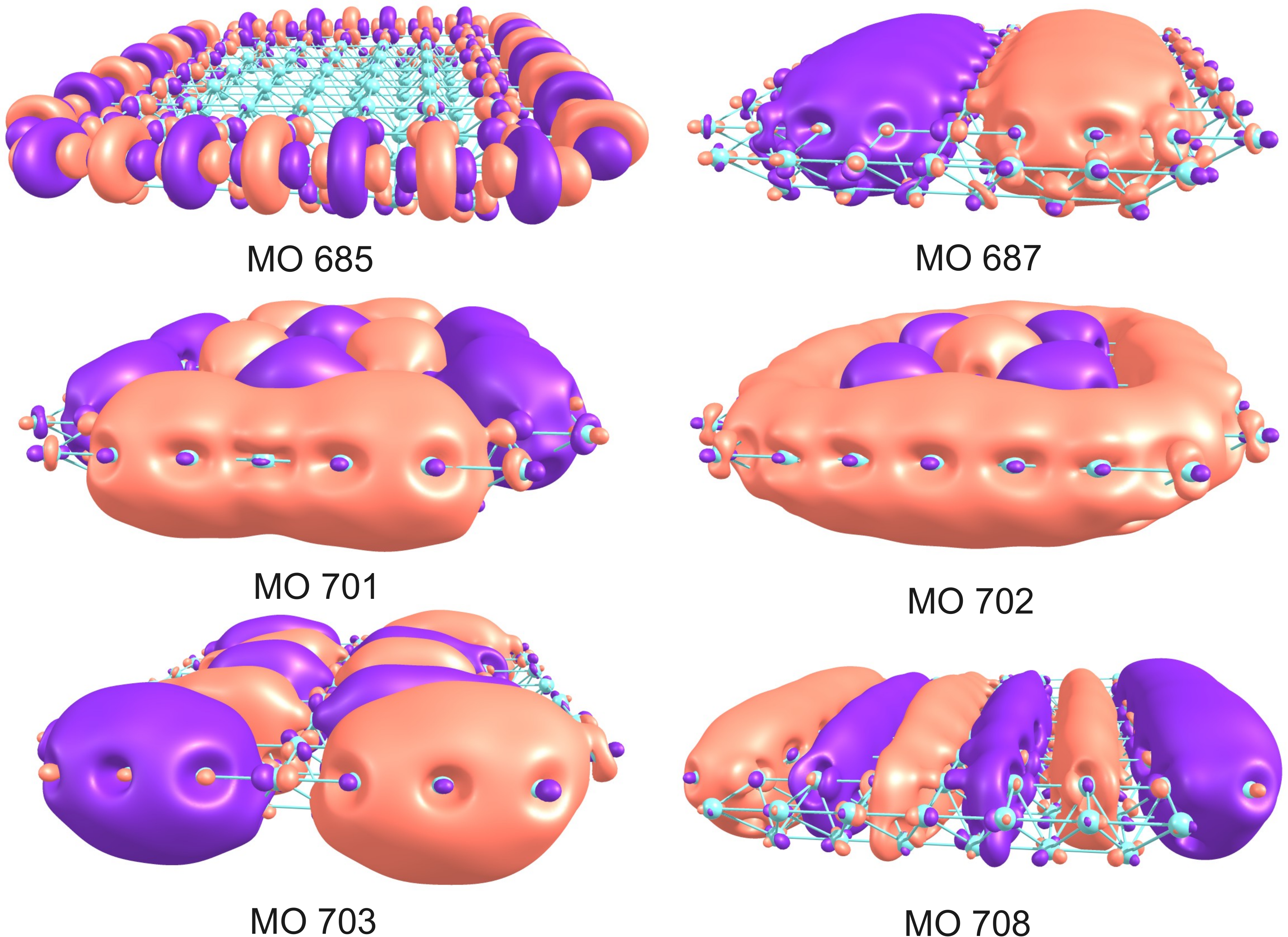}
	\caption{Some characteristic MOs of the bare Cu(001) calculation.}
	\label{fig:surface-MOs}
\end{figure}

\begin{figure}[t]
\includegraphics[width=\columnwidth]{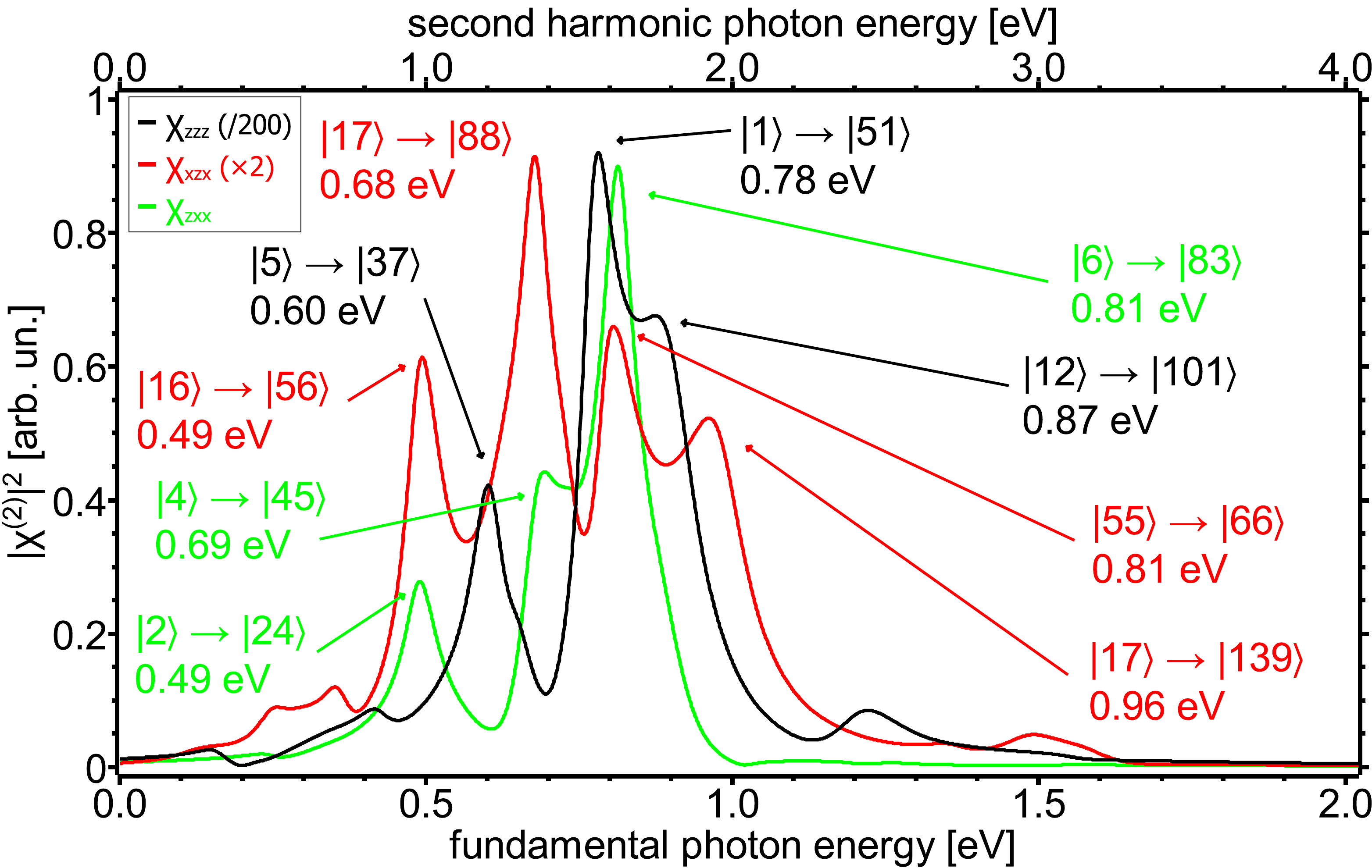}
\includegraphics[width=\columnwidth]{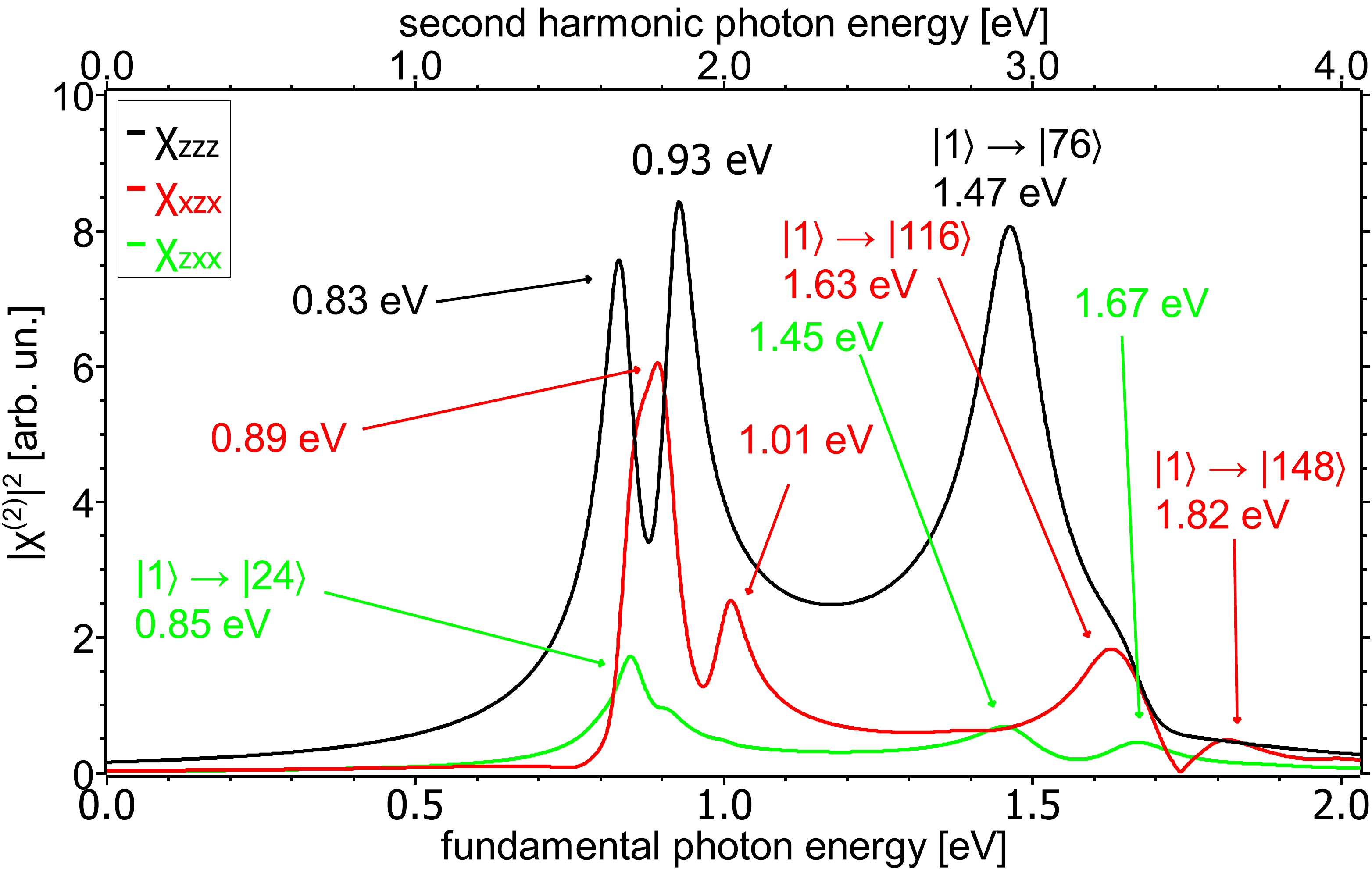}
\caption{Theoretically calculated non-zero elements of the second harmonic susceptibility tensor, $\chi^{(2)}_{zxx}$, $\chi^{(2)}_{xzx}$, and $\chi^{(2)}_{zzz}$  for FeOEP deposited on Cu(001) (top), and the bare Cu(001) surface (bottom). Wherever possible the pertinent many-body-state transitions are given as well. The spectra are calculated at room temperature $T=300$ K. Please be aware of the magnitude difference of the elements of FeEOP at the surface. }
    \label{fig:theoretical-FeOEP-surface-SHG}
\end{figure}

Generally, charge transfer plays an important role in most of the transitions in the FeOEP/Cu(001) system. For example, in  $\chi^{(2)}_{\mathrm{zzz}}$ the peak at 0.781~eV is due to the many-body-state transition $\vert 1 \rangle\!\rightarrow\!\vert 51 \rangle$. The ground state $\vert 1 \rangle$ consists of the virtual transitions from MO 912 to MO 920 (with amplitude 0.045), 
from MO 917 to MO 921 (with amplitude 0.083), 
from MO 918 to MO 922 (with amplitude 0.081), 
from MO 915 to MO 923 (with amplitude 0.034), 
from MO 912 to MO 920 (with amplitude 0.055),  
and from MO 917 to MO 925 (with amplitude 0.055). 
The excited many-body state   $\vert 51 \rangle$ consists of the virtual excitations 
from MO 918 to MO 921 (with amplitude 0.5261), 
from MO 917 to MO 922 (with amplitude 0.343), 
from MO 917 to MO 926 (with amplitude 0.134),  
and from MO 918 to MO 925 (with amplitude 0.224).  Inspection of all MOs involved in the excitation $\vert 1 \rangle\!\rightarrow\!\vert 51 \rangle$ (through the reduced one-electron transition density  matrix) reveals that this peak corresponds to roughly half an electron getting transferred from FeOEP into the surface, i.e.\ this is mainly a charge-transfer peak. 
Please note that the strongly localized many-body states can be loosely thought of as corresponding to the $\Gamma$ point of a calculation in reciprocal space (e.g.\ density function theory), especially in the case of low surface coverage (since little-to-negligible overlap between neighboring porphyrin molecules leads to rather dispersionless states). Fig.\ \ref{fig:surface-MOs} depicts some MOs of the bare Cu(001) surface. One can see different types of orbitals, such as edge states (MO 685), localized $p$- and $d$-character states (MOs 687, 701, and 702), and standing-wave-like states (MOs 703 and 708).

\begin{table}
	\caption{Localization of charges for different total charges and multiplicities of the combined FeOEP-surface system at the Hartree-Fock level.}
	\begin{tabular}{cSSS}
		\hline\hline
		 multiplicity &  \multicolumn{1}{c}{\begin{tabular}{c}total\\charge\end{tabular}}  &  \multicolumn{1}{c}{\begin{tabular}{c}FeEOP\\charge\end{tabular}} & \multicolumn{1}{c}{\begin{tabular}{c}substrate\\charge\end{tabular} }\\ 
		\hline
		1 & 0 & 0.746 & -0.746\\
		1 & 2 & 0.822 & 1.177\\
		2 & 1 & 0.809 & 0.191\\
		2 & 3 & 1.052 & 1.948\\
		3 & 0 & 0.763 & -0.763\\
		3 & 2 & 0.831 & 1.169\\
		4 & 1 & 0.810 & 0.191\\
		4 & 3 & 2.133 & 0.867\\
		\hline\hline
	\end{tabular}
	\label{tab:charge-diffusion}
\end{table}

In order to better understand the influence of the charge-transfer states on the localized charges of FeOEP, we also perform a series of calculations in which we vary the total number of electrons in the whole system as well as the total multiplicity. The results show that for higher total charges, the additional charge gets diffused into the surface, leaving the FeEOP system with a typical charge of 1 electron (Tab.~\ref{tab:charge-diffusion}). The only exception is the combination of high multiplicity (quartet) and high total charge (+3), which however can be easily explained since it leads to an energetically high electronic state. It is also a known fact that for high multiplicities a restricted HF calculation (i.e.\ a calculation in which the spatial part of the spin-up and spin-down electrons is kept identical) yield higher energies due to the occupational imbalance of the two electron species (this however is usually remedied by the post-HF methods, in our case the CC method).

In summary, our approach allowed to theoretically compute the energy spectrum of both FeOEP and FeOEP deposited on a Cu(100) surface. In this context, we identify one-electron virtual excitations, which fall into three main categories: FeOEP-substrate interactions, states localized on the iron-porphyrin molecule, and pure surface states. A fourth category, namely states localized at the edges of the substrate (Cu bilayer), which could be artifacts of the specific geometry chosen, do not contribute to the virtual excitations of the many-body states. Furthermore, we identify charge-transfer states between the iron-porphyrin and the Cu(100) substrate, which are responsible for the second-harmonic-generation signal. We show that the surface absorbs any additional electronic charge, so that the iron-porphyrin entity has a net charge closely corresponding to one electron removed. For a real system, these charge-transfer states are relatively stable since the charge gets diffused even further away into the (quasi-)infinitely extended substrate \cite{OTERO2017105}.

\section{Comparison and Discussion}
\label{compar}

\begin{figure}[t]
\includegraphics[width=0.8\columnwidth]{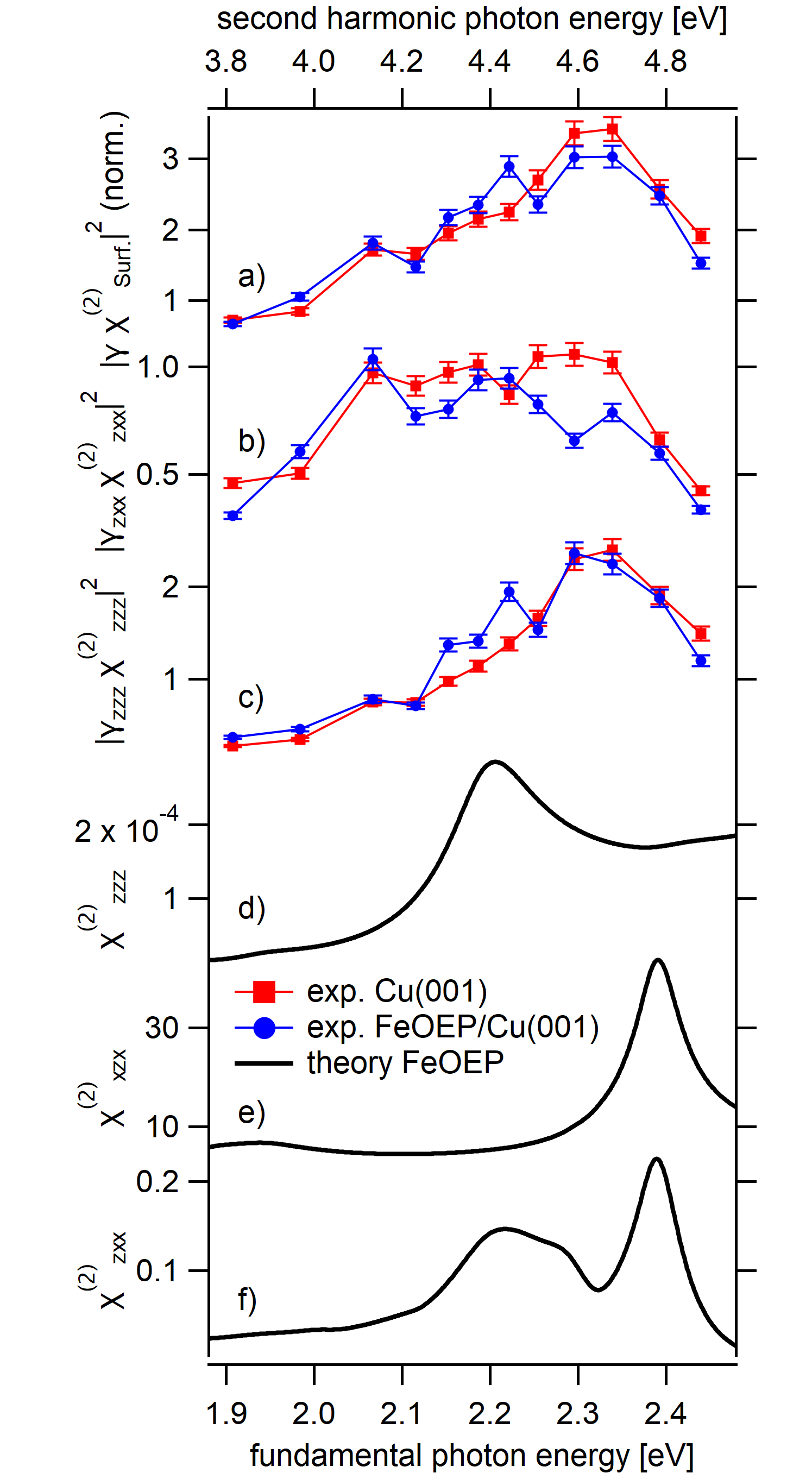}
\caption{Measured SHG spectra for the bare Cu(001) substrate (red squares) and the FeOEP/Cu(001) interface (blue circles), namely a) $|\gamma\chi^{(2)}_{\mathrm{surf.}}|^2$, which is acquired with p-P polarization combination,  b) $|\gamma\chi^{(2)}_{\mathrm{zxx}}|^2$ acquired with s-P polarization combination, and c) $|\gamma\chi^{(2)}_{\mathrm{zzz}}|^2$, which represents the difference between these two signals, followed by the calculated non-linear susceptibility tensor components d) $\chi^{(2)}_{\mathrm{zzz}}$, e) $\chi^{(2)}_{\mathrm{xzx}}$ and f) $\chi^{(2)}_{\mathrm{zxx}}$ for the FeOEP molecule (black lines), in dependence on the fundamental respectively second harmonic photon energy. }
    \label{fig3}
\end{figure}

For a comparison with the calculated second harmonic susceptibility tensor elements $\chi_{ijk}^{(2)}$, we measure the SH yield for fundamental photon energies $\hbar\omega$ ranging from 1.9~eV to 2.5~eV and both p-P and s-P polarization combinations. The resulting SHG spectra, normalized to the intensity of the fundamental beam as described in section~\ref{exp_setup}, are displayed in Fig.~\ref{fig3}. The topmost curve (Fig.~\ref{fig3}a) shows the SHG signal acquired in the p-P polarization combination for 1 ML FeOEP/Cu(001) and Cu(001). For both samples, it exhibits a continuous increase from $\hbar\omega = 1.9$~eV to about 2.35~eV, followed by a decrease until the end of the measured fundamental photon energy range at 2.5~eV. Compared to Cu(001), the presence of the FeOEP molecular adsorbate leads to an increase of the SHG signal in the 2.1-2.25~eV range and a suppression at higher photon energies. The s-P polarization combination (Fig.~\ref{fig3}b) shows a less pronounced photon energy dependence overall, with FeOEP/Cu(001) leading to a decrease of the SHG signal compared to Cu(001) at $\hbar\omega \geq 2.1$~eV, which is particularly strong near 2.15~eV and 2.3~eV. 

We then further analyze our experimental data in order to separate the different $\chi_{ijk}^{(2)}$ components, which contribute to the respective second harmonic fields as described in section~\ref{exp_pol}. While the s-P polarized SHG signal is proportional to $|\gamma_{zxx}\chi_{zxx}^{(2)}|^2$, the p-P signal contains three non-linear susceptibility tensor elements, namely $\chi_{zzz}^{(2)}$, $\chi_{zxx}^{(2)}$ and $\chi_{xzx}^{(2)}$. Since we found that the S-polarized SHG yield, which contains $\chi_{xzx}^{(2)}$, is about one order of magnitude smaller than the p-P polarized yield [compare Fig.~\ref{fig2}(b-c)], we neglect the contribution of this tensor element. Consequently, the expression for the p-P SHG intensity can then be approximated as: 
\begin{eqnarray}
	I^{(2)}_{\text{p}-\text{P}} &\approx& |\gamma_{zzz}\chi_{zzz}^{(2)} + \gamma_{zxx}\chi_{zxx}^{(2)}|^2, \\
	I^{(2)}_{\text{p}-\text{P}} &\approx& |\gamma_{zzz}\chi_{zzz}^{(2)}|^2 + |\gamma_{zxx}\chi_{zxx}^{(2)}|^2 + \nonumber\\ &&+|\gamma_{zzz}\gamma_{zxx}\chi_{zzz}^{(2)}\chi_{zxx}^{(2)}|.
	\label{shintensities}
\end{eqnarray}

The effect of the molecular adsorbate on the linear coefficients $\gamma_{zzz}$ and $\gamma_{zxx}$ is expected to be well below 1\%, i.e.\ negligible \cite{Dvorak2000}. We thus estimate the absolute values of these linear coefficients for Cu(001): Following the respective literature \cite{Sipe1987}, we derive that $|\gamma_{zzz}|$ is one order of magnitude smaller than $|\gamma_{zxx}|$, while both share the same spectral dependence with a maximum near $\hbar\omega=2.3$~eV due to a resonance when the photon energy is approaching the Cu $3d$ band \cite{Jepsen1981, Courths1984, Petrocelli1993}. The maximum at $\approx 2.35$~eV in the p-P SHG yield shown in Fig.~\ref{fig3}a) is thus caused by a resonant enhancement of the SHG process in the Cu substrate at $1\omega$, similar to a previous observations of $2\omega$ resonance at higher photon energies \cite{Lupke1994, Bisio2009}. $|\gamma_{zzz}\gamma_{zxx}|$ is again one order of magnitude smaller than $|\gamma_{zzz}|$. This estimate now allows us to neglect the cross-term $|\gamma_{zzz}\gamma_{zxx}\chi_{zzz}^{(2)}\chi_{zxx}^{(2)}|^2$ and consequently to approximate the $zzz$ second harmonic contribution by
\begin{equation}
	|\gamma_{zzz}\chi_{zzz}^{(2)}|^2 \approx I^{(2)}_{\text{p}-\text{P}} - I^{(2)}_{\text{s}-\text{P}}.
\end{equation}

The result of the subtraction is shown in Fig.~\ref{fig3}c). We clearly identify an enhancement of the surface SHG response of about 30~\% by the FeOEP adsorbate in $|\gamma_{zzz}\chi_{zzz}^{(2)}|^2$ at around $\hbar\omega=2.2$~eV. A comparison with the calculated non-linear susceptibility tensor elements (see Fig.~\ref{fig3}d-f) further strengthens our assignment of this SHG feature to FeOEP, as the calculated $\chi^{(2)}_{zzz}$ shows a maximum at around $\hbar\omega=2.2$~eV as well (see Fig.~\ref{fig3}d). The experimentally determined $\chi^{(2)}_{zxx}$ (Fig.~\ref{fig3}b) also exhibits peaks at $\hbar\omega=2.25$~eV and slightly below $\hbar\omega=2.4$~eV, i.e. near the calculated peak positions of $\chi^{(2)}_{\mathrm{zxx}}$ (Fig.~\ref{fig3}f). However, its identification in the experimental SHG spectra appears to be overshadowed by the underlying strong signature of the $1\omega$ resonance of the Cu(001) substrate in this photon energy range, as discussed above, and no enhancement of the SHG response is observed here. Slight differences of the peak positions, in particular for $\chi^{(2)}_{\mathrm{zxx}}$ near $\hbar\omega=2.35$~eV, might be due to some remaining influence of the Cu(001) substrate, which is not included in the calculations in this photon energy range due to computational limitations. 

The overall very good agreement of the experimental SHG spectra of the FeOEP/Cu(001) surface with calculated SHG spectra of FeOEP reinforces that our approach is successful in identifying molecular transitions in the SHG spectra. Our quantum chemical approach thus makes it possible to assign specific molecular transitions to particular spectral features. In particular, the many-body excitations $|2\rangle\!\rightarrow\!|14\rangle$ and $|4\rangle\!\rightarrow\!|17\rangle$ are reflected in $\chi_{zzz}^{(2)}$ (compare Fig.~\ref{fig:theoretical-SHG}). The agreement of the experimentally observed spectral features with calculated spectra of isolated FeOEP moreover shows that the interaction with the Cu(001) substrate indeed mostly leads to an intensity enhancement of the spectral features of FeOEP, as discussed above. Substrate-induced spectral shifts (see Fig.~\ref{fig:theoretical-FeOEP-surface-SHG}) in contrast take place in the lower photon energy range outside of the experimentally covered range. 

\begin{figure}[t]
{   \centering
    \includegraphics[width=\columnwidth]{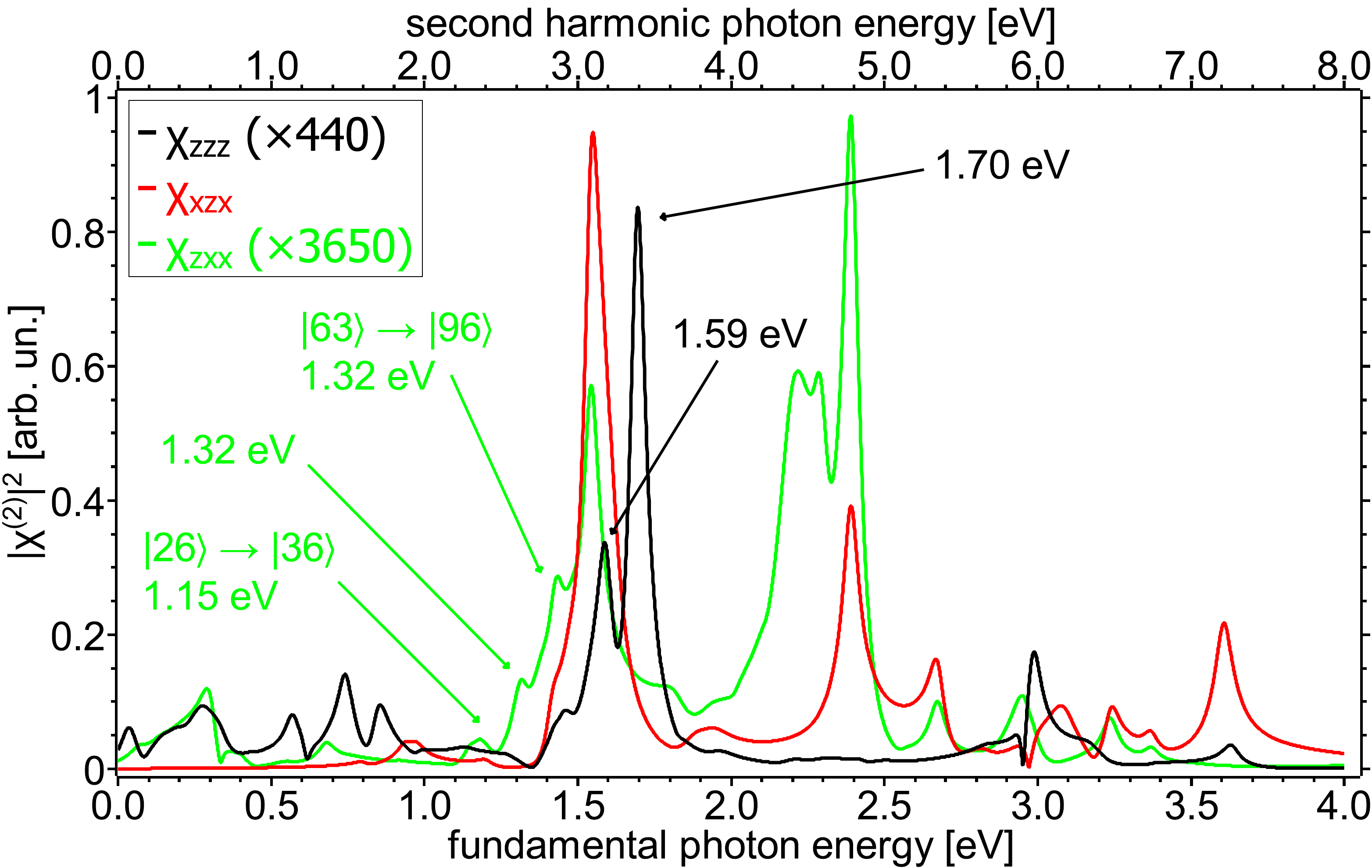}
    \caption{Theoretically calculated non-zero elements of the second harmonic susceptibility tensor after pumping (see text), $\chi^{(2)}_{zzz}$ (440 times magnified), $\chi^{(2)}_{xzx}$, and $\chi^{(2)}_{zxx}$ (3650 times magnified) for the bare FeOEP molecule in arbitrary units. Indicated are some peaks which are not present without pumping (cf.~Fig.~\ref{fig:theoretical-SHG}).
    The spectra are calculated at room temperature $T=300$ K before pumping and have an effective electronic temperature $T_{\text{eff}}\approx 2085$ K after pumping.}
    \label{fig:pump-SHG}
}
\end{figure}

The experimental SHG spectra in Fig.~\ref{fig3} appear to be slightly broader than the calculated ones, especially at higher photon energies, which might be an effect of the ultrashort experimental pulse duration and the corresponding spectral width. Besides the temperature, which is identical in experiment and theory, the pump laser itself can also actively alter the SHG spectrum \cite{PRB-chi3}. In order to check possible effects of the employed few 10~fs laser pulses on the SHG response, we recalculate the spectra but introduce a new distribution $f_\text{pump}$ which is a superposition of the thermalized Boltzmann distribution $f$ and a small proportion of the linear spectrum, since we assume that the pump pulse can populate only states which are optically directly addressable from the ground state, which is exactly what $\chi^{(1)}$ describes
\begin{eqnarray}
    f_\text{pump}(E_a)=(1-\alpha)f(E_a)+\alpha \frac{\chi^{(1)}(E_a)}{\sum_j\chi^{(1)}(E_j)}.
\end{eqnarray}
For our purposes the factor $\alpha$ is set to 0.01 to render the effect small but visible. The denominator of the second term normalizes $\chi^{(1)}$ so that $\int_{-\infty}^{\infty}\chi^{(1)}(E)\,dE=1$. Starting from the expectation value of the total energy for a specific $T$ in a thermalized system
\begin{eqnarray}
    \langle E_\text{tot}(T)\rangle=\sum_j f(E_j) E_j,
\end{eqnarray}
we can then define the effective temperature $T_\text{eff}$ after the pump pulse as the temperature for which the total energy of the nonequilibrated system after the pump pulse matches a thermalized state
\begin{eqnarray}
    \langle E_\text{tot}(T_\text{eff})\rangle=\sum_j f_\text{pump}(E_j) E_j.
\end{eqnarray}
We find $T_\text{eff}\approx 2085$ K (only electronic temperature). Note, that because the heat capacity of the phonons is much higher (due to the much larger mass of the nuclei), the temperature after complete thermalization turns out to be only a few tenths of K. Fig.\ \ref{fig:pump-SHG} shows the $\chi^{(2)}_{zzz}$ before and after pumping. We see that some new peaks appear at 1.169 and 1.544 eV due to the optically triggered population of higher excited states $|26\rangle$ and $|63\rangle$. Due to the optical selection rules the tensor elements $\chi^{(2)}_{xzx}$ and $\chi^{(2)}_{zxx}$ remain almost unaffected (not shown here). Interestingly, this effect can also introduce further asymmetry in the system (since now the Hamiltonian must reflect the spatial
symmetry of the atom arrangement 
in combination with any external fields \cite{PhysRev.136.B864}). 

Compared to earlier work on thicker molecular films, namely phtalocyanines or porphyrins prepared on insulating or semiconducting substrates \cite{Kumagai1993, Hoshi1995, Hoshi1996, Yamada1996, Echevarria2003, Pandey2016}, we here extend the detection of molecular spectral features down to the single monolayer range and a metallic substrate. The direct comparison with SHG spectra derived from first principles calculations moreover provides an increased understanding of the sensitivity of SHG to specific molecular transitions compared to more commonly employed simpler models of the non-linear optical response. Moreover, molecular resonances for $\hbar\omega \geq 2$~eV can be clearly identified, and their enhancement understood due to the charge-transfer character of the molecule-substrate interaction, by means of our quantum chemical approach.

\section{Conclusions\label{summary}}

From a polarization- and energy-dependent analysis of the SHG yield of 1~ML FeOEP/Cu(001), we were able to derive the modification of the SHG spectrum with respect to the bare Cu(001) surface by the molecular adsorbate. Specifically, a molecule-induced enhancement of SHG in the 2.15-2.35~eV photon energy range could be linked to the influence of the $\chi_{zzz}^{(2)}$ non-linear susceptibility tensor element. 
First-principles calculations of the molecular electronic states and SHG spectra reveal that this is due to many-body excitations $|2\rangle\!\rightarrow\!|14\rangle$ and $|4\rangle\!\rightarrow\!|17\rangle$. 
We clearly show that the interaction with the substrate exhibits a strong charge-transfer character and thus leads to an enhancement of the excitations along the z direction, in particular the $\chi^{(2)}_{zzz}$ tensor element. This, in turn, is reflected in the SHG response both at higher energies, as revealed by the observation of an enhancement of $\chi^{(2)}_{zzz}$ at energies resonant with molecular transitions in the experiment, and at lower photon energies, as indicated by the appearance of additional features in the theoretical SHG spectra. 

We believe that our results help elucidate the interaction mechanisms between molecular magnets and substrates, and can thus facilitate the active manipulation of their electronic and magnetic properties, potentially by ultrafast optical excitation. Future pump-probe SHG spectroscopy experiments might be able to directly address the resulting property changes in the time domain. 

\section*{Data Availability Statement}

The data are available upon reasonable request from the authors. 

\begin{acknowledgments}

We thank Eberhard Riedle, Jens G\"udde and Heiko Wende for valuable experimental advice. Funded by the Deutsche Forschungsgemeinschaft (DFG, German Research Foundation) - Project-ID 278162697 - SFB 1242. Further financial support from the DFG through SPP 1840 QUTIF (project ES 492/1) is gratefully acknowledged.

\end{acknowledgments}


{
\section*{Appendix A: Character tables of some relevant double point groups}
Tables \ref{tab:characterTableC4v} and \ref{tab:characterTableD4h} show the character tables of the double point groups C$^*_{\text{4v}}$ and D$_{\text{4h}}^*$, which represent the exact and the approximate symmetry of the FeOEP, as discussed in the text.

A doublet state (only spin) in  C$^*_{\text{4v}}$ and D$_{\text{4h}}^*$ belongs to the $\Gamma_6$ and the $\Gamma_6^+$ irreps, respectively. A quartet state (only spin) has the irreps $\Gamma_6\oplus\Gamma_7$ and $\Gamma_6^+\oplus\Gamma_7^+$, respectively (we follow the convention that the parity of the electron is $+1$). 

\begin{table*}[t]
{
\caption{Character table of the double point group C$^*_{\text{4v}}$. The symmetry operations with the overbar correspond to the normal symmetry operations followed by a 360$^\circ$ spatial rotation around any arbitrary axis. The table is presented in the form, in which the upper left part corresponds to the simple C$_{\text{4v}}$ point group, for which the Schoenflies notation of the irreps are given. $\Gamma_7$ has also the basis $\Big(\left|\frac32,\frac32\rangle\right.,\left|\frac32,-\frac32\rangle\right.\Big)$.\label{tab:characterTableC4v}}
\begin{tabular}{c||ccccc|cc|cc}
    C$^*_{\text{4v}}$ & $E$ & $C_2$ & 2$C_4$ & 2$\sigma_{\text{v}}$ & 2$\sigma_{\text{d}}$ & $\bar{E}$ &2$\bar{C}_4$&
    standard basis\\
    &&$\bar{C}_2$&&2$\bar{\sigma}_{\text{v}}$ & 2$\bar{\sigma}_{\text{d}}$&&
    \\\hline\hline 
    A$_1$ ($\Gamma_1$) & 1 & 1 & 1 & 1 & 1 & 1 & 1 &$z$, $x^2+y^2$, $z^2$\\
    A$_2$ ($\Gamma_2$) & 1 & 1 & 1 & $-1$ & $-1$ & 1 & 1 &$I_z$\\
    B$_1$ ($\Gamma_3$) & 1 & 1 & $-1$ & 1 & $-1$ & 1 & $-1$ &$x^2-y^2$\\
    B$_2$ ($\Gamma_4$) & 1 & 1 & $-1$ & $-1$ & 1 & 1 & $-1$ &$xy$\\
    E ($\Gamma_5$) & 2 & $-2$ & 0 & 0 & 0 & 2 & 0 & $(x,y)$, $(I_x,I_y)$, $(xy,xz)$
    \\\hline
    $\Gamma_6$ & 2 & 0 & $\sqrt{2}$ & 0 & 0 & $-2$ & $-\sqrt{2}$
    &$\Big(\left|\frac12,\frac12\rangle\right.,\left|\frac12,-\frac12\rangle\right.\Big)$\\
    $\Gamma_7$ & 2 & 0 & $-\sqrt{2}$ & 0 & 0 & $-2$ & $\sqrt{2}$
    &$\Big((x^2-y^2)\left|\frac12,\frac12\rangle\right.,(x^2-y^2)\left|\frac12,-\frac12\rangle\right.\Big)$\\
\end{tabular}
}
\end{table*}

\begin{table*}[t]
{
\caption{Character table of the double point group D$^*_{\text{4h}}$. The symmetry operations with the overbar correspond to the normal symmetry operations followed by a 360$^\circ$ spatial rotation around any arbitrary axis. The table is presented in the form, in which the upper left part corresponds to the simple D$_{\text{4h}}$ point group, for which the Schoenflies notation of the irreps are given. $\Gamma_7^+$ has also the basis $\Big(\left|\frac32,\frac32\rangle\right.,\left|\frac32,-\frac32\rangle\right.\Big)$.\label{tab:characterTableD4h}}
\begin{tabular}{c||cccccccccc|cccc|cc}
 D$^*_{\text{4h}}$ & $E$ & 2$C_4$ & 2$C_2$ & 2$C'_2$ & 2$C''_2$ & $i$ & 2$S_4$ & $\sigma_\text{h}$ & 2$\sigma_\text{v}$ & 2$\sigma_\text{d}$ & $\bar{E}$ & 2$\bar{C}_4$ & $\bar{i}$ & $\bar{S}_4$ &
    standard basis\\ 
    &&&2$\bar{C}_2$ & 2$\bar{C}'_2$ & 2$\bar{C}''_2$ & && $\bar{\sigma}_\text{h}$ &2$\bar{\sigma}_\text{v}$ & 2$\bar{\sigma}_\text{d}$ &&&& \\\hline\hline
    A$_{1\text{g}}$ ($\Gamma_1^+$) & 1 & 1 & 1 & 1 & 1 & 1 & 1 & 1 & 1 & 1 & 1 & 1 & 1 & 1 & $x^2+y^2$, $z^2$ \\
    A$_{2\text{g}}$ ($\Gamma_2^+$) & 1 & 1 & 1 & $-1$ & $-1$ & 1 & 1 & 1 & $-1$ & $-1$ & 1 & 1 & 1 & 1 & $I_z$ \\
    B$_{1\text{g}}$ ($\Gamma_3^+$) & 1 & $-1$ & 1 & 1 & $-1$ & 1 & $-1$ & 1 & $-1$ & 1 & 1 & $-1$ & 1 & $-1$ &  $x^2-y^2$\\
    B$_{2\text{g}}$ ($\Gamma_4^+$) & 1 & $-1$ & 1 & $-1$ & 1 & 1 & $-1$ & 1 & $-1$ & 1 & 1 & $-1$ & 1 & $-1$ &  $xy$\\
    E$_{\text{g}}$ ($\Gamma_5^+$)& 2 & 0 & $-2$ & 0 & 0 & 2 & 0 & $-2$ &0 &0 & 2 &0 &2 & 0 & $(I_x,I_y)$, $(xz,yz)$ \\
    A$_{1\text{u}}$ ($\Gamma_1^-$) & 1 & 1 & 1 & 1 & 1 & $-1$ & $-1$ & $-1$ & $-1$ & $-1$ & 1 & 1 & $-1$ & $-1$ &  $(x^2-y^2)xyz$\\
    A$_{2\text{u}}$ ($\Gamma_2^-$) & 1 & 1 & 1 & $-1$ & $-1$ & $-1$ & $-1$ & $-1$ & 1 & 1 & 1 & 1 & $-1$ & $-1$ &  $z$\\
    B$_{1\text{u}}$ ($\Gamma_3^-$) & 1 & $-1$ & 1 & 1 & $-1$ & $-1$ & 1 & $-1$ & 1 & $-1$ & 1 & $-1$ & $-1$ & 1 &  $xyz$\\
    B$_{2\text{u}}$ ($\Gamma_4^-$) & 1 & $-1$ & 1 & $-1$ & 1 & $-1$ & 1 & $-1$ & 1 & $-1$ & 1 & $-1$ & $-1$ & 1 &  $(x^2-y^2)z$\\
    E$_{\text{u}}$ ($\Gamma_5^-$)& 2 & 0 & $-2$ & 0 & 0 & $-2$ & 0 & 2 &0 &0 & 2 &0 &$-2$ & 0 & $(x,z)$ \\\hline
    $\Gamma_6^+$ & 2 & $\sqrt{2}$ & 0 & 0 & 0 & 2 & $\sqrt{2}$ & 0 & 0 & 0 &$-2$ & $-\sqrt{2}$ &$-2$ & $-\sqrt{2}$ &$\Big(\left|\frac12,\frac12\rangle\right.,\left|\frac12,-\frac12\rangle\right.\Big)$\\
    $\Gamma_7^+$ & 2 & $-\sqrt{2}$ & 0 & 0 & 0 & 2 & $-\sqrt{2}$ & 0 & 0 & 0 &$-2$ & $\sqrt{2}$ &$-2$ & $\sqrt{2}$ &$\Big(I_y\left|\frac12,\frac12\rangle\right.,I_x\left|\frac12,-\frac12\rangle\right.\Big)$\\
    $\Gamma_6^-$ & 2 & $\sqrt{2}$ & 0 & 0 & 0 & $-2$ & $-\sqrt{2}$ & 0 & 0 & 0 &$-2$ & $-\sqrt{2}$ &$2$ & $\sqrt{2}$ &$\Big((x^2-y^2)xyz\left|\frac12,(x^2-y^2)xyz\frac12\rangle\right.,\left|\frac12,-\frac12\rangle\right.\Big)$\\
    $\Gamma_7^-$ & 2 & $-\sqrt{2}$ & 0 & 0 & 0 & $-2$ & $\sqrt{2}$ & 0 & 0 & 0 &$-2$ & $\sqrt{2}$ &2 & $-\sqrt{2}$ &$\Big(y\left|\frac12,\frac12\rangle\right.,x\left|\frac12,-\frac12\rangle\right.\Big)$\\
\end{tabular}
}
\end{table*}
}


\clearpage


\end{document}